\documentclass[twocolumn]{aastex631}

\usepackage{amsmath}
\usepackage{bm}
\usepackage{tabularx}
\usepackage{cancel}

\newcommand{\ZX}[2]{#2}  % for arXiv
\submitjournal{ApJS}

\shorttitle{PULSAR-SMBH TIMING MODEL}
\shortauthors{Zexin Hu, et al.}
\begin{document}

%----------------------------------------------------------------------
\title{A Realistic Pulsar -- Supermassive Black Hole Timing Model}

\author[0000-0002-3081-0659]{Zexin Hu}
\affiliation{Department of Astronomy, School of Physics, Peking University, 
Beijing 100871, China}
\affiliation{Kavli Institute for Astronomy and Astrophysics, Peking University, 
Beijing 100871, China}

\author[0000-0002-8742-8397]{Ziming Wang}
\affiliation{Department of Astronomy, School of Physics, Peking University, 
Beijing 100871, China}
\affiliation{Kavli Institute for Astronomy and Astrophysics, Peking University, 
Beijing 100871, China}

\author[0000-0002-1334-8853]{Lijing Shao}
\affiliation{Kavli Institute for Astronomy and Astrophysics, Peking University, 
Beijing 100871, China}
\affiliation{National Astronomical Observatories, Chinese Academy of Sciences, 
Beijing 100101, China}

\correspondingauthor{Lijing Shao}
\email{lshao@pku.edu.cn}
%----------------------------------------------------------------------



\begin{abstract}
Timing observation of pulsars orbiting around a supermassive black hole (SMBH)
can measure the spacetime around the SMBH to a high precision and thus be a
novel probe of the gravity theory. Future high-frequency surveys of the Galactic
Centre (GC) region to be performed by the next-generation radio telescopes, such
as the SKA, may discover pulsars that orbit around Sagittarius~A* (Sgr~A*), the
SMBH dwelling in our GC. In this paper, we present a realistic pulsar-SMBH
timing model based on the post-Newtonian equations of motion of the pulsar.
Considering the expected timing precision in the future, we take into account
several next-to-leading order light propagation time delays in the timing model.
For the first time, we include the effects of proper motion of Sgr~A*, which
were expected to break the spin measurement degeneracy. We forecast the
measurement precision of various parameters of Sgr~A*, and discuss the data
analysis procedure in the presence of red noise, which can be strong if the
pulsar is a normal pulsar. The realistic timing model constructed in this study
will serve as a useful tool in future searching and timing of pulsar-SMBH
systems in the GC.
\end{abstract}


%----------------------------------------------------------------------


%----------------------------------------------------------------------
\section{Introduction} 
\label{sec:intro}
%----------------------------------------------------------------------

Black holes (BHs) are one of the most fascinating objects predicted by General
Relativity (GR), and measuring their fundamental properties can provide various
tests of gravity theory in the strong-field region. Observations have revealed
the existence of at least two types of astrophysical BHs, namely the
stellar-mass BHs and the supermassive BHs (SMBHs). Since the first detection of
gravitation waves (GWs) from a binary BH merger by the Advanced LIGO in
2015~\citep{LIGOScientific:2016aoc}, GW observations have provided hundreds of
measurements of stellar-mass BHs~\citep{LIGOScientific:2025hdt,
LIGOScientific:2025slb} and brought the tests of GR to high post-Newtonian (PN)
orders~\citep{LIGOScientific:2016lio}. On the other hand, although it is
believed that SMBHs exist at the center of most massive
galaxies~\citep{McConnell:2012hz, Kormendy:2013dxa}, measuring their properties
accurately remains a challenge. Current measurements of the spin of SMBHs mainly
rely on the spectroscopy~\citep{Reynolds:2013qqa}. But this method suffers from
strong selection biases which favour high-spin BHs~\citep{Reynolds:2019uxi}, and
can be unreliable for sources accreting near or above the Eddington
rate~\citep{Riaz:2019kat}. 

The SMBH in \ZX{}{the centre of our Milky Way Galaxy (or Galactic 
Centre, hereafter referred to as GC)}, Sagittarius~A* (Sgr~A*), is the closest SMBH
to us. Monitoring the stars orbiting around \ZX{}{the GC} has provided a precise
measurement of the mass of the SMBH, which is around $4.3\times 10^6\,M_\odot$,
and its distance to the Solar System of about $8\,{\rm
kpc}$~\citep{GRAVITY:2020gka}. The detection of the Schwarzschild precession of
the S2 orbit shows the appealing potential of testing gravity theory with
Sgr~A*. However, stars closer to Sgr~A* need to be found for measuring the spin
of the SMBH with astrometric observations~\citep{Waisberg:2018qzl}. \ZX{}{Moreover,} the
measurements come from the Event Horizon Telescope (EHT) observations of Sgr~A*
currently can only loosely suggest a moderate to large spin
magnitude~\citep{EventHorizonTelescope:2022wkp,EventHorizonTelescope:2024rju}. 

Timing pulsars orbiting around Sgr~A* can be a novel probe of the SMBH
properties~\citep{Wex:1998wt, Pfahl:2003tf, Liu:2011ae, Psaltis:2015uza,
Zhang:2017qbb, Hu:2023ubk, Hu:2024blq, Shao:2025vmb}, as well as theories of
gravity~\citep{Dong:2022zvh,Hu:2023vsg}. Previous studies have shown that timing
one pulsar with an orbital period less than $\sim0.5\,{\rm yr}$ or combining the
timing of two pulsars with orbital periods less than $\sim 5\,{\rm yr}$ can both
measure the spin of Sgr~A* to better than $1\%$ level~\citep{Liu:2011ae,
Hu:2024blq}. Further, timing a pulsar with a tight orbit can even measure the
quadrupole moment of Sgr~A*~\citep{Wex:1998wt}, which provides a test of the
no-hair theorem in GR~\citep{Carter:1971zc}.

Current surveys have not found any close pulsars orbiting around Sgr~A* yet. In
fact, there are only several pulsars that have been found in the GC
region~\citep{Deneva:2009mx, Johnston:2006fx, Eatough:2013nva, Rea:2013pqa,
Lower:2024sdi, SKAOPulsarScienceWorkingGroup:2025syv}, much fewer than the
expected population of pulsars there. The ``missing pulsar problem'' may be
partially caused by the complex interstellar medium (ISM) environment in the GC
region, which makes the detection a challenge for observations at the \ZX{}{typically
low frequencies used to observe pulsars}~\citep{Cordes:2002wz}, while high-frequency surveys at the
current stage are limited by the steep spectrum of the pulsar emission and the
instrument sensitivity. Astrophysical explanation considering the formation and
evolution of pulsars at the GC may also account for the absence of
pulsars~\citep{Dexter:2013xga}. Nevertheless, future observations with
next-generation radio telescopes \ZX{}{will provide the best means} to find pulsars in the GC
and even close pulsars orbiting around Sgr~A*~\citep{Liu:2011ae,Bower:2018mta,
Desvignes:2025hkk}.

The upcoming opportunities of finding and timing pulsars orbiting around Sgr~A*
demand the development of a realistic timing model that can be used in real data
analysis for \ZX{}{pulsar-SMBH systems}. The so-called timing model
describes the relation between the times of arrival (TOAs) of pulsar pulses
observed at the Earth telescopes and the proper rotation of the pulsar in its
own inertial frame~\citep{Damour:1986, Hu:2023vsq}.  Due to the orbital motion
of the pulsar as well as the curved spacetime caused by the SMBH, the pulses
observed on Earth are not simply periodic signals. A comparison between the
model prediction and observed TOAs will enable the measurement of system
parameters.

The timing model for the pulsar-SMBH system is significantly different from
those models used for currently timed binary pulsars, mainly due to the grand
mass of the SMBH. The large orbital semi-major axis of the pulsar in the
pulsar-SMBH system leads to leading-order time delays at the order of
$10^4\,{\rm s}$ for a pulsar with an orbital period $P_b \sim 0.5\,{\rm
yr}$~\citep{Hu:2023ubk}. The large time delays enable the detection of
next-to-leading-order (NLO) effects in this system. Besides, the angular
momentum of the system is dominated by the spin of the SMBH, thus the spin-orbit
coupling is important in this system~\citep{Wex:1998wt}. Many high-order effects
that just begin to be measured in the Double Pulsar
system~\citep{Kramer:2021jcw} should be easily detectable in a pulsar-SMBH
system and therefore need to be taken into account for a realistic timing model.

Currently, the development of the pulsar-SMBH timing model is mainly based on
the analytic solution of the second-order PN motion of compact binary systems
with spin, derived by~\citet{Wex:1995}, and incorporation of the quadrupole
effects of the SMBH perturbatively~\citep{Liu:2011ae, Psaltis:2015uza}. The
analytic timing model takes the key advantage of a parametrized approach, which
can be partially theory independent~\citep{Damour:1991rd}. It is also
computationally cheap compared to, for example, a full GR model~\citep[see
e.g.,][]{Zhang:2017qbb}. However, the analytical timing model is relatively hard
to describe additional effects come from various perturbations such as the
astrophysical environment around Sgr~A* \citep[see e.g.,][for perturbations from
dark matter]{Hu:2023ubk} or corrections from modified gravity theories
\citep{Dong:2022zvh, Hu:2023vsg}, especially for those periodic effects that
deform the pulsar orbit from an ellipse.

In this paper, we aim to provide a comprehensive description of a realistic
numerical timing model based on previous developments in~\citet{Hu:2023ubk}. We
here further complete the timing model by incorporating all detectable
high-order effects, so that the timing model provided in this paper can be
applied to real data analysis. We also give a detailed discussion on the
parameter estimation procedure based on the numerical timing model and mock data
simulations. Particularly, we study the parameter estimation in the presence of
red noise, which can be important in the pulsar-SMBH system for a normal pulsar
(instead of a millisecond pulsar). This is usually expected after taking  the
scattering effects of pulsar pulses with the interstellar medium  into
consideration \citep{SKAOPulsarScienceWorkingGroup:2025syv}. 

The remaining part of this paper is organized as follows. In
Section~\ref{sec:timing model} we describe the details of the realistic
pulsar-SMBH timing model. In Section~\ref{sec:parameter estimation}, we forecast
the expected measurement precision of the SMBH properties for future
observation. We discuss the data analysis procedure in the presence of red noise
in Section~\ref{sec:red noise}. Section~\ref{sec:conclusions} summarizes our
results.

%----------------------------------------------------------------------
\section{Timing model} \label{sec:timing model}
%----------------------------------------------------------------------

A pulsar timing model describes the relation between the observed TOAs of pulsar
pulses and the stable proper rotation of the pulsar in its inertial
frame~\citep{Damour:1986}, which is an essential part for timing data analysis.
In this section, we focus only on the timing model for the pulsar-SMBH system,
which describes the TOAs at the Solar System Barycenter (SSB) and of infinite
frequency. We first describe the propagation of the radio signal in the curved
SMBH spacetime in Section~\ref{subsec:light propagation}, followed by the
orbital dynamics of the pulsar in Section~\ref{subsec:orbital dynamics}. A
complete description of the timing model is shown in Section~\ref{subsec:timing
model}. In Section~\ref{subsec:inverse timing model}, we construct an efficient
inverse timing model for real data analysis.  As an overview, we summarize
various effects considered in the pulsar orbital motion in
Table~\ref{tab:effects of different terms}, and give a fiducial case study of
pulsar orbits and time delays in Figure~\ref{fig:orbit} and Figure~\ref{fig:time
delay}, respectively.

%----------------------------------------------------------------------
\subsection{Light propagation} \label{subsec:light propagation}
%----------------------------------------------------------------------

A radio pulse signal emitted by a pulsar needs to travel through the curved
spacetime of the SMBH before it arrives at the telescopes on Earth. In the
pulsar timing procedure, various time delays are used to describe this process.
At the leading order, the time for light to travel through the flat spacetime is
accounted by the so-called R\"{o}mer delay
\begin{equation}\label{eq:Romer delay}
	\Delta_{\rm R}=\frac{1}{c}\bm{r}\cdot\hat{\bm{K}}_0=\frac{z}{c}\,,
\end{equation}
where $\bm{r}$ is the relative position vector pointing from the SMBH to the
pulsar, $\hat{\bm{K}}_0$ is the unit vector of the line of sight direction, and
$z \equiv \bm{r}\cdot\hat{\bm{K}}_0$. A time shift corresponding to the distance
$d$ between the Sgr~A* and SSB is subtracted from the expression, which, in the
constant case, is only a matter of redefinition of the reference time epoch. The
effects caused by the relative motion between the SSB and Sgr~A* are discussed
in Section~\ref{subsec:timing model}.

The lowest-order propagation time delay caused by the spacetime curvature of the
pulsar companion is also well studied and can be measured in many binary pulsar
systems~\citep{Manchester:2004bp, Freire:2024adf}. It is described by the 1PN
Shapiro delay~\citep{Shapiro:1964uw,Blandford:1976},
\begin{equation}\label{eq:1PN Shapiro}
	\Delta_{{\rm S}1}=-\frac{2GM}{c^3}\ln\left(r-z\right)\,,
\end{equation}
where $M$ is the mass of the SMBH and $r=|\bm{r}|$. A constant term is again
subtracted from the expression, assuming that the distance between Sgr~A* and
SSB tends to infinity, which is reasonable when considering the PN effects. One
should notice that the splitting of R\"{o}mer delay and Shapiro delay as above
is a coordinate-dependent concept~\citep{Damour:1991rd}, and in this paper, we
always work in the harmonic coordinates centered on the SMBH. 

Although in current pulsar timing observations, the NLO effects caused by higher
PN corrections to the light propagation have only started to be observed in a
limited number of systems, for example, in the Double Pulsar
system~\citep{Kramer:2021jcw}, a rough estimation suggests that the 2PN Shapiro
delay should be observable for the pulsar-SMBH system. For a pulsar in a
circular orbit with an orbital period $P_b=1\,{\rm yr}$ and a moderate orbital
inclination $i=30^\circ$, the amplitude of the 1PN Shapiro delay can be
estimated by Equation~(\ref{eq:1PN Shapiro}), 
\begin{equation}
	\mbox{amplitude of } \Delta_{{\rm S}1} \sim\frac{2GM}{c^3}\ln\frac{1+\sin i}{1-\sin i}\sim
	50\,{\rm s}\,.
\end{equation}
The 2PN Shapiro delay should roughly have an amplitude at the order of
$\beta_O^2\times 50\,{\rm s}\sim 10\,{\rm ms}$, where the characteristic orbital
velocity parameter is
\begin{equation}
	\beta_O=\frac{(GMn_b)^{1/3}}{c}\,,
\end{equation}
with $n_b \equiv 2\pi/P_b$~\citep{Damour:1991rd}. The exact amplitude of the 2PN
effects can be even larger for orbits with large eccentricity and inclination
angles closer to $90^\circ$.

The expected timing precision of GC pulsars for future next-generation
telescopes such as the SKA is expected to reach at least $\sigma_{\rm
TOA}=1\,{\rm ms}$~\citep{Liu:2011ae}. Thus, the 2PN Shapiro delay should be
taken into account in the pulsar-SMBH timing model. The analytic expression of
the 2PN Shapiro delay in the harmonic coordinates is given
by~\citet{Klioner:2010pu},
\begin{equation}
	\Delta_{\rm{S}2}=\Delta_{\rm \Delta PN}+\Delta_{\rm PPN}\,,
\end{equation}
where
\begin{equation}
	\Delta_{\rm \Delta
	PN}=-\frac{4G^2M^2}{c^5r}\left(1-\frac{z}{r}\right)^{-1}\,,
\end{equation}
and
\begin{equation}
	\Delta_{\rm PPN}=-\frac{G^2M^2}{4c^5r}\left(\frac{z}{r}-15
	\frac{\arccos\left(-z/r\right)}{\sqrt{1-z^2/r^2}}\right)\,,
\end{equation}
with the range of the arccosine function $[0,\pi]$. In these expressions, we
have again taken the distance between the Sgr~A* and SSB to infinity.

For the Double Pulsar system, the 2PN Shapiro delay is taken into account in an
elegant way, which can be written as 
\begin{equation}
	\Delta_{\rm S}=\Delta_{\rm S1}+\Delta_{\rm S2}=-2{r_{\rm S}}\ln(\Lambda_u+\delta 
	\Lambda_u^{\rm len})\,.
\end{equation}
Here, \ZX{}{$r_{\rm S}$} is the so-called range parameter \ZX{}{commonly used in the Shapiro delay}. 
The detailed definition of \ZX{}{$r_{\rm S}$}, $\Lambda_u$ and
$\delta\Lambda_u^{\rm len}$ can be found in~\citet{Kramer:2021jcw}. This
simplified expression is based on the fact that the term $\Delta_{\rm PPN}$ is
bounded as~\citep{Klioner:2010pu}
\begin{equation}
	|\Delta_{\rm PPN}|\leq\frac{15}{4}\pi\frac{G^2M^2}{c^5d_i}\,,
\end{equation} 
where $d_i$ is the impact parameter for the light travel in a straight line. For
the Double Pulsar system, taking $d_i$ as the radius of a neutron star, and $M$
to be at the order of one Solar mass, this gives an upper bound $\sim 20\,{\rm
\mu s}$, which is below the timing precision and can be safely
ignored~\citep{Kramer:2021jcw}. However, for the pulsar-SMBH system, this bound
is not small enough as it gives a value as large as the 1PN effect. Our
simulations also show that $\Delta_{\rm PPN}$ can be the same order of, or even
larger than, $\Delta_{\rm \Delta PN}$ for the pulsar-SMBH system, so that we
need to take both of them into account. 

A simple estimation of the amplitude of the 2PN Shapiro delay and the future
conservatively-estimated timing precision for GC pulsars in fact suggests that
Shapiro delay at even higher PN orders might be detectable for a pulsar-SMBH
system with a small orbital period, large orbital eccentricity and nearly
edge-on inclination, though the probability for finding such pulsars might be
rather low for future observations. Here we stop at the 2PN level, and
higher-order contributions, if necessary, can be added in a similar manner.

The above discussions have not considered the spin of the SMBH. The Shapiro
delay describes the additional light propagation time in the Schwarzschild
spacetime, while the spin of the SMBH leads to the so-called frame-dragging
effect~\citep{Lense:1918zz,Wex:1998wt}. For the pulsar-SMBH system, the
leading-order time delay caused by the frame-dragging effect is effectively at
the 2PN order and is~\citep{Wex:1998wt},
\begin{equation}
	\Delta_{\rm FD}=-\chi\frac{2G^2M^2}{c^5}\frac{\hat{\bm{s}}\cdot
	\left(\hat{\bm{K}}_0\times\hat{\bm{n}}\right)}{r-z}\,,
\end{equation}
where $\hat{\bm{n}} \equiv \bm{r}/r$, $\hat{\bm{s}} \equiv \bm{S}/S$, and
$\chi=cS/GM^2$ is the dimensionless spin of the SMBH.

Now we have listed out all the time delays corresponding to light propagation in
the curved spacetime of the SMBH, consistent to the 2PN order. Higher-order
effects can be added similarly if necessary. We should mention that analytic
solutions for null geodesics in the Schwarzschild or Kerr spacetime can be
expressed with various kinds of elliptic functions and parametrized by the
constants of motion~\citep{Hackmann:2018fmk,Cieslik:2023qdc}. One can then
numerically solve the so-called emitter-observer problem to get the time delays
for light propagation. This procedure automatically takes into account full GR
effects. However, we choose to use the expanded formulas introduced in this
subsection, as they are computationally cheaper, easier to extend for
astrophysical perturbations or gravity theories beyond GR, and they are very
convenient for the construction of the inverse timing model in
Section~\ref{subsec:inverse timing model}. 

Finally, as mentioned at the beginning, we ignore the various effects caused by
the ISM, such as dispersion or scattering \citep{Lorimer:2005misc}. These
effects are expected to be frequency-dependent and might be removed with
multi-frequency observation.

%----------------------------------------------------------------------
\subsection{Orbital dynamics}
\label{subsec:orbital dynamics}
%----------------------------------------------------------------------

For our purpose to build a timing model that can be applied in future data
analysis, we first treat the pulsar as a test particle moving in the spacetime
of the central SMBH. We use the 2PN equations of motion and keep the
leading-order effects of the spin-orbital coupling and the SMBH quadrupole
moment \citep{Hu:2023ubk}. The equations of motion of the pulsar in the harmonic
coordinates read,
\begin{eqnarray}\label{eq:equation of motion}
	\ddot{\bm{r}}=\ddot{\bm{r}}_{\rm N}+\ddot{\bm{r}}_{\rm 1PN}
	+\ddot{\bm{r}}_{\rm SO}+\ddot{\bm{r}}_{\rm Q}+\ddot{\bm{r}}_{\rm 2PN}\,,
\end{eqnarray}
where dots denote the derivative with respect to the coordinate time $t$.
Besides the Newtonian acceleration, $\ddot{\bm{r}}_{\rm N}=-GM\hat{\bm{n}}/r^2$,
other terms in Equation~(\ref{eq:equation of motion}) represent the 1PN and 2PN
corrections and the leading-order contributions from the SMBH spin and
quadrupole moment. Their expressions are~\citep{Barker:1975ae,Damour:1988mr},
\begin{subequations}\label{eq:details acceleration}
\begin{eqnarray}
	\ddot{\bm{r}}_{\rm 1PN}&=&-\frac{GM}{c^2r^2}\left[\left(-\frac{4GM}{r}+
	v^2\right)\hat{\bm{n}}-4\dot{r}\bm{v}\right]\,,\\
	\ddot{\bm{r}}_{\rm SO}&=&\chi\frac{6G^2M^2}{c^3r^3}\bigg[\hat{\bm{s}}\cdot
	\left(\hat{\bm{n}}\times\bm{v}\right)\hat{\bm{n}}+\dot{r}\left(\hat{\bm{n}}
	\times\hat{\bm{s}}\right)\nonumber\label{eq:SO}\\
  && \hspace{1.7cm} -\frac{2}{3}\left(\bm{v}\times\hat{\bm{s}}\right)
	\bigg]\,,\\
	\ddot{\bm{r}}_{\rm Q}&=&-q\frac{3G^3M^3}{2c^4r^4}\bigg\{\left[5\left(
	\hat{\bm{n}}\cdot\hat{\bm{s}}\right)^2-1\right]\hat{\bm{n}}
  \nonumber\\
  && \hspace{1.7cm} -2\left(\hat{
	\bm{n}}\cdot\hat{\bm{s}}\right)\hat{\bm{s}}\bigg\}\,,\\
	\ddot{\bm{r}}_{\rm 2PN}&=&\frac{G^2M^2}{c^4r^3}\left[\left(2\dot{r}^2-\frac{
	9GM}{r}\right)\hat{\bm{n}}-2\dot{r}\bm{v}\right]\,,
\end{eqnarray}
\end{subequations}
where $\bm{v}=\dot{\bm{r}}$ and $v=|\bm{v}|$. Similar to the spin, $q \equiv
c^4Q/G^2M^3$ is the dimensionless quadrupole moment of the SMBH, with $Q$ the
(dimensionful) quadrupole moment. By writing down the above equations, we have
made several assumptions and simplifications to the orbital motion of the
pulsar, and we justify these assumptions as follows.

In above equations, we have assumed that the spacetime of the SMBH is
axisymmetric, like the Kerr spacetime. But we do not impose the constraint from
the no-hair theorem, which states that the spacetime of a Kerr BH is fully
determined by the BH mass and spin, and thus the dimensionless quadrupole moment
$q$ depends on the dimensionless spin parameter $\chi$ via
$q=-\chi^2$~\citep{Thorne:1980ru}. The independent treatment of $\chi$ and $q$
is especially convenient when one is aiming to test the no-hair theorem, as this
timing model can provide an independent measurement of the two parameters. 

In Equations~(\ref{eq:equation of motion}) and (\ref{eq:details acceleration}),
we have neglected the finite mass ratio of the pulsar and SMBH, which is at the
order of $m_{\rm PSR}/M\lesssim 10^{-6}$. At the Newtonian order, this only
leads to an effective (mis)identification of the SMBH mass $M$ to  be $M+m_{\rm
PSR}$. Even though the theoretical estimation of the SMBH mass measurement
precision in such a pulsar-SMBH system can reach a relative error of $10^{-6}$
or better for a pulsar with a very small orbital period, $P_b\lesssim 0.1\,{\rm
yr}$~\citep{Liu:2011ae,Hu:2023ubk}, it seems unrealistic to have such a precise
measurement of the SMBH mass in practice  due to the complex astrophysical
environment around Sgr~A*. This simple argument then suggests that the effects
caused by the finite mass ratio are likely to be unobservable in the near
future, considering the probability of finding such a close pulsar and the lack
of detailed modeling of other environmental perturbations \citep{Hu:2026aez}. Therefore, we ignore
all corrections corresponding to the small mass ratio.

For the term $\ddot{\bm{r}}_{\rm SO}$ shown in Equation~(\ref{eq:SO}), we have
not specified a spin supplementary condition (SSC), which corresponds to the
ambiguity of choosing the representation point for spinning objects in
GR~\citep{Barker:1974}. Though the detailed expression of the spin-orbit
coupling term depends on the choice of SSC, they only introduce a fractional
change at the order of the mass ratio $m_{\rm PSR}/M$ for compact objects like
neutron stars~\citep{Kidder:1995zr,Li:2024cvj}, and thus it can be ignored in
our calculation. From another estimation, different SSCs will introduce a shift
of the representation point of the pulsar at the order of $vS_{\rm PSR}/m_{\rm
PSR}c^2$~\citep{Barker:1974}, which gives a time delay at the order of
$0.1\,{\rm \mu s}$ for a millisecond pulsar with an orbital period $\sim 1\,{\rm
yr}$. This is far below the expected timing precision and thus is negligible.
Also,  we ignore the spin-orbit coupling term contributed by the pulsar's spin
and the spin-spin coupling, as the spin of pulsar in the system is much smaller
than that of the SMBH~\citep{Wex:1998wt}.

We truncate the expansion of equations of motion~(\ref{eq:equation of motion})
at the 2PN level, as the higher-order contributions can be shown to be
negligible for the expected timing precision of the SKA and realistic
observation time spans. The higher-order PN corrections can be separated into
two parts: one is the conservative part starting from the 3PN level, which is
similar to what we have included in the current equations, and the other one is
the back reaction of the GW radiation starting from the 2.5PN level and it is
proportional to the small mass ratio. For the 3PN corrections, it will deform
the orbit, leading to a periodic effect at the order of $\beta_O^6\times
\Delta_{\rm R}\sim 1\,{\rm \mu s}\,(P_b/1\,{\rm yr})^{-4/3}$, which is too small
to be detected. At the same time, it will also cause an additional periastron
advance with $\dot{\omega}^{\rm 3PN}\sim \beta_O^4\dot{\omega}^{\rm 1PN}$, where
$\omega$ is the longitude of the periastron and $\dot{\omega}^{\rm
1PN}=3GMn_b/c^2 a (1-e^2)$ with $a$ and $e$ the semi-major axis and eccentricity
of the pulsar orbit respectively~\citep{Damour:1986}. Though the orbital
eccentricity may further amplify $\dot{\omega}^{\rm 3PN}$, for orbits with
moderate eccentricity \ZX{}{($e\lesssim 0.8$)}, we may roughly account for the effects of eccentricity to
a factor of $10$ and can estimate the time delay caused by the secular effect to
be
\begin{equation}
	\dot{\omega}^{\rm 3PN} T_{\rm obs} a\sin i/c\sim 1\,{\rm ms}\, \left(
	\frac{T_{\rm obs}}{5\,{\rm yr}} \right) \left(\frac{P_b}{1\,{\rm
	yr}}\right)^{-7/3} \sin i\,,
\end{equation}
with $T_{\rm obs}$ the total observing time span. Having said that, similar to
the higher-order terms of the Shapiro delay, the estimation in fact suggests
that for pulsar-SMBH systems with small $P_b$, large $e$, and a nearly edge-on
inclination, also with a large $T_{\rm obs}$ or a better timing precision, the
secular effect of 3PN corrections might be observable. However, a similar
argument as for ignoring the mass ratio also applies to the 3PN corrections, as
the secular effects caused by PN corrections are largely degenerate with the
secular effects caused by the environmental perturbations, and it seems
impractical to model the environmental effects to such a high precision. This is
different when considering secular effects, as they can have distinctive
signatures and be separated.

The dissipative 2.5PN correction will cause a decay of the pulsar orbit, which
provided the first evidence for the existence of GWs in the binary pulsar system
PSR~B1913$+$16~\citep{Taylor:1979zz}. The leading-order change rate of the
pulsar orbital period caused by GW emission can be expressed
as~\citep{Peters:1963ux, poisson_will_2014},
\begin{equation}
	\dot{P}_b=-\frac{192\pi}{5} \eta_{\rm sym} f(e)\beta_O^5\,,
\end{equation}
with 
\begin{equation}
	f(e)=(1-e^2)^{-7/2}\left(1+\frac{73}{24}e^2+\frac{37}{96}e^4\right)\, ,
\end{equation}
\ZX{}{and $\eta_{\rm sym}=m_{\rm PSR}M/(m_{\rm PSR}+M)^2\approx m_{\rm PSR}/M$.}
For a pulsar in a pulsar-SMBH system with an orbital period $P_b=1\,{\rm yr}$ and
orbital eccentricity $e=0.8$, one has $\dot{P}_b\sim-5\times10^{-12}$. This
leads to a cumulated time delay at the order of 
\begin{equation}
	 \pi \dot{P}_b\frac{T^2}{P_b^2}\frac{a\sin i}{c}\sim30\,{\rm \mu s}
	 \left(\frac{T_{\rm obs}}{5\,{\rm yr}}\right)^2\left(\frac{P_b}{1\,{\rm yr}}
	 \right)^{-3}\sin i\,.
\end{equation}
Therefore, for a total observation time span at the order of $10\,{\rm yr}$, it
is not necessary to consider the orbital decay caused by GWs, while such an
observation time span already permits a precise measurement of the SMBH spin and
quadrupole moment~\citep{Liu:2011ae,Hu:2023ubk}. 

Finally, as mentioned before, various environmental perturbations also affect
the orbital motion of the pulsar. For example, \citet{Merritt:2009ex} studied
the effects of stellar mass objects in the context of using stellar orbits to
measure the SMBH properties. \citet{Hu:2023ubk} have studied the effects caused
by dark matter spike around the SMBH. It is interesting and important to study
how to model the environmental effects in our GC and how they will affect the
timing observations, though it is beyond the scope of this work and we leave it
for future study.

%----------------------------------------------------------------------
\begin{figure}[t]
	\begin{center}
		\includegraphics[trim=150pt 0pt 0pt 0pt,clip,width=8cm]{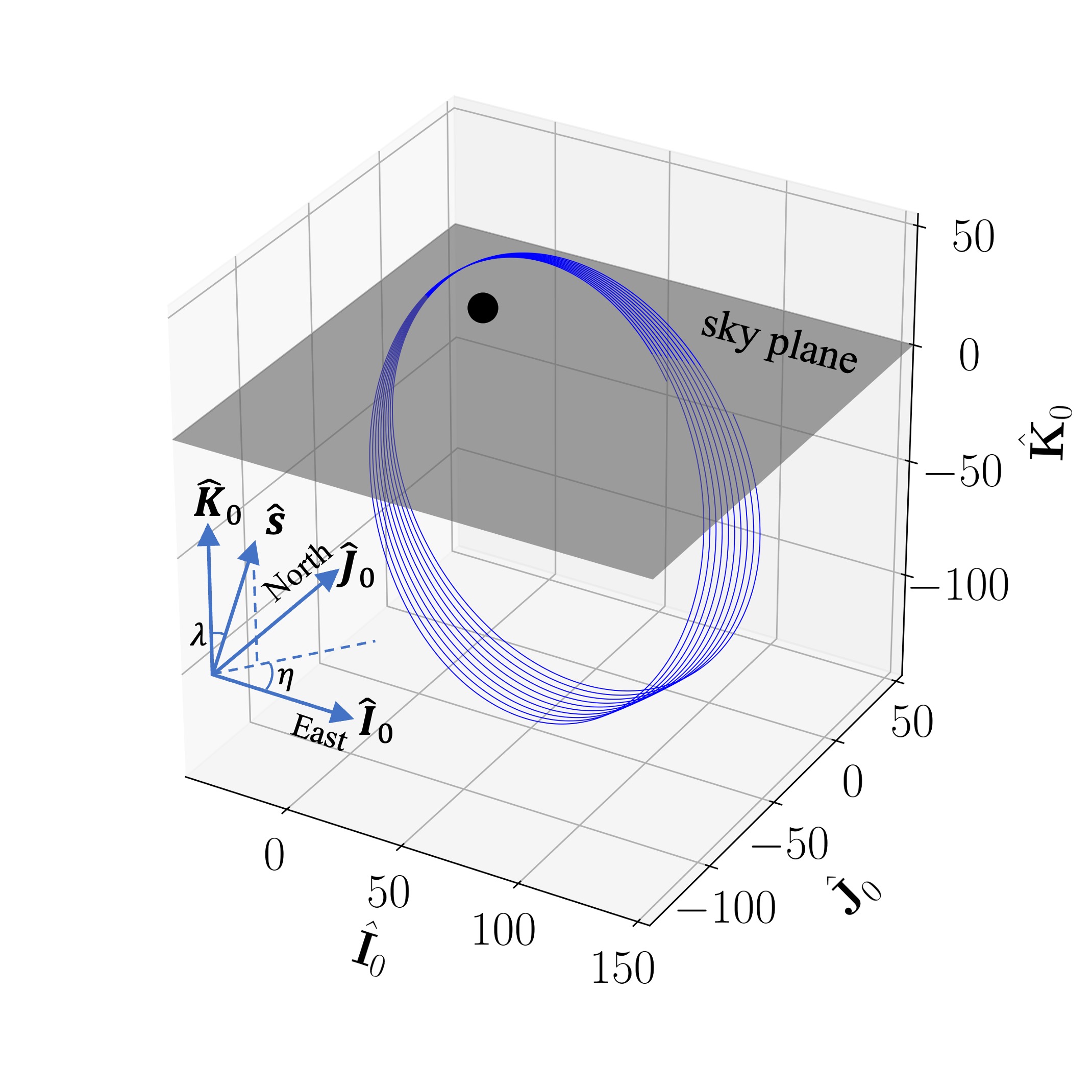}
	\caption{Illustration of the pulsar orbit with system parameters given in
	Equation~(\ref{eq:system params}). The axes are in the unit of AU. \label{fig:orbit}}
	\end{center}
\end{figure}
%----------------------------------------------------------------------

The above discussions justify the validity of using Equations~(\ref{eq:equation
of motion}) and (\ref{eq:details acceleration}) to describe the pulsar orbital
motion for our timing model. As a fiducial case, in Figure~\ref{fig:orbit}, we
illustrate the orbit of a pulsar for a total time span $T_{\rm obs}=5\,{\rm yr}$
in a system with the following parameters,
\begin{subequations}\label{eq:system params}
	\begin{eqnarray}
		M&=&4.3\times10^6\,M_\odot\,,\\
		\chi&=&0.6\,,  \hspace{0.3cm} \lambda=\frac{\pi}{6}\,, 
		\hspace{0.3cm} \eta=\frac{5\pi}{9}\,, \hspace{0.3cm}q=-0.36\,,\\
		P_b&=&0.5\,{\rm yr}\,, \hspace{0.3cm} e=0.8\,,\\
		i&=&\frac{\pi}{5}\,, \hspace{0.3cm} \omega=\frac{5\pi}{7}\,, 
		\hspace{0.3cm} \Omega=0\,, \hspace{0.3cm}
		f_0=-\frac{3\pi}{4}\,,
	\end{eqnarray}	
\end{subequations}
where $\lambda$ and $\eta$ describe the direction of the SMBH spin relative to
the coordinate frame $(\hat{\bm{I}}_0, \hat{\bm{J}}_0, \hat{\bm{K}}_0)$ with
$\hat{\bm{I}}_0$, $\hat{\bm{J}}_0$ pointing towards east and north respectively.
Then we have $\hat{\bm{s}}=(\sin\lambda\cos\eta, \sin\lambda\sin\eta,
\cos\lambda)$, as shown in the figure. Parameters in $\Theta_{\rm
orb}=\left\{P_b,e,i,\omega,\Omega,f_0\right\}$ are the usual orbital elements
describing a Keplerian orbit~\citep{poisson_will_2014}, and some of them have
been defined before except that $\Omega$ is the longitude of the ascending node
and $f_0$ is the initial orbital true anomaly. The values of the SMBH parameters
$\Theta_{\rm SMBH} =\left\{M,\chi, \lambda,\eta, q\right\}$ are inspired by
results from~\citet{GRAVITY:2020gka} and~\citet{EventHorizonTelescope:2024rju},
where the former gives an accurate measurement of the SMBH mass and the latter
suggests a moderate to large spin with a small inclination.\footnote{Note that
there is a sign difference in the definition of the line of sight.} Here, we use
a moderate value for the dimensionless spin, so that it may represent a
conservative case for estimating the spin measurement precision. The angles for
the orbital parameters are randomly chosen to avoid special orientations that
might render the parameter-estimation problem degenerate. Because of the PN
correction, the orbit of the pulsar in fact is not a Keplerian orbit. Thus, one
should regard the orbital parameters in the sense of osculating elements of the
orbit at a reference time. They are equivalent to the initial position and
velocity vectors of the pulsar~\citep{poisson_will_2014}. 

%%%
\begin{table*}[t]
	\centering
	\caption{Approximate time delays related to different effects in the
	pulsar's orbital motion.  The second column gives an order-of-magnitude
	estimation for the time delays caused by different effects, while the last
	column shows if the effect is included in our timing model. Except for the
	Newtonian term related to the Keplerian orbit, we estimate the cumulative
	time delays caused by secular effects. The third column shows the estimation
	results for the fiducial system with system parameters  in
	equations~(\ref{eq:system params}). For terms including the mass ratio
	$m_{\rm PSR}/M$, $m_{\rm PSR}$ is taken to be $1.4\,M_\odot$, and for the
	SSC term we take the rotation period of the pulsar to be $1\,{\rm ms}$.
	Dimensionless $P_{b1} \equiv P_b /(1\,{\rm yr})$ and $T_{\rm obs1} \equiv
	T_{\rm obs}  /(1\,{\rm yr})$ are respectively $P_b$ and $T_{\rm obs}$  in
	unit of year. Note
	that the dependence on orbital parameters in the second column is not complete and can only serve
	as a reference.}
	\label{tab:effects of different terms}
	\renewcommand{\arraystretch}{1.6}
	\begin{tabularx}{\textwidth}{@{\extracolsep{\fill}} llcc}
		\hline\hline
		Effects & Time Delay [s] & Time Delay (fiducial) [s] & Included\\
		\hline
		Newtonian & $8\times 10^4\,P_{b1}^{2/3}\sin i$
		& $3\times 10^4$ & Yes\\
		1PN & $4\times 10^2\,P_{b1}^{-1}T_{{\rm obs1}}
		(1-e^2)^{-1}\sin i$& $7\times 10^3$ & Yes\\
		Spin-orbit coupling& $4\,\chi P_{b1}^{-4/3}T_{\rm obs1}(1-e^2)^{-3/2}\sin i$ 
		& $80$ & Yes\\
		2PN & $0.7\,P_{b1}^{-5/3}T_{{\rm obs1}}(1-e^2)^{-2}
		\sin i$ & $50$ & Yes\\
		Quadrupole & $0.03\,q P_{b1}^{-5/3}T_{\rm obs1}(1-e^2)^{-2}\sin i$  
		&$1$ & Yes\\
		Mass ratio & $3 \times 10^{-2} \,P_{b1}^{2/3}\sin i$ &$1\times 10^{-2}$ & No\\
		Radiation reaction & $6\times 10^{-9} \,P_{b1}^{-3}T_{\rm obs1}^2(1-e^2)^{-7/2}f(e)
		\sin i$ & $1\times 10^{-4}$ & No \\
		Higher PN & $2 \times 10^{-4} \,P_{b1}^{-7/3}T_{\rm obs1}\sin i$ & $3 \times 10^{-3}$ & No\\
		SSC & $2 \times 10^{-7} \,P_{b1}^{-1/3}$ & $2 \times 10^{-7}$ & No\\
		\hline
	\end{tabularx}
\end{table*}
%%%

Before ending this subsection, we want to discuss the spin evolution of the
pulsar. Due to the curved spacetime of the SMBH, the rotation axis of the pulsar
suffers a precession. At the leading order, the precession of the spin of the
pulsar can be written as~\citep{Barker:1979}
\begin{equation}
	\frac{{\rm d}\hat{\bm{e}}_3}{{\rm d}t}=(\bm{\Omega}_{\rm DS}+
	\bm{\Omega}_{\rm LT})\times\hat{\bm{e}}_3\,,
\end{equation}
where $\hat{\bm{e}}_3$ is the direction of the spin axis of the pulsar in the
center of mass frame, the one where we describe the pulsar's orbital motion. 
\ZX{}{Geodetic} precession and Lense-Thirring precession are 
\begin{eqnarray}
	\bm{\Omega}_{\rm DS}&=&\frac{3}{2}\frac{GM}{c^2r^3}\bm{r}\times\bm{v}\,,\\
	\bm{\Omega}_{\rm LT}&=&\chi\frac{G^2M^2}{c^3r^3}
	\left[3(\hat{\bm{n}}\cdot\hat{\bm{s}})
	\hat{\bm{n}}-\hat{\bm{s}}\right]\,,
\end{eqnarray}
respectively.  For a pulsar in a circular, $P_b = 1\,{\rm yr}$ orbit, the
precession rate caused by them are $|\bm{\Omega}_{\rm DS}|\sim0.1\,{\rm deg\
yr^{-1}}$ and $|\bm{\Omega}_{\rm LT}|\sim10^{-3}\,{\rm deg\ yr^{-1}}$. The
precession rate in the pulsar-SMBH system is rather low compared to the rate in
compact binary pulsar systems~\ZX{}{\citep{Kramer:2021jcw, Fonseca:2025rqd}} as the spacetime curvature
here is in fact smaller. Such a small precession rate seems sub-dominant in near
future observations, and we take $\hat{\bm{e}}_3$ as a constant direction in our
analysis.

As a short summary and for useful references, we conclude all the effects
discussed in Table~\ref{tab:effects of different terms}.

%----------------------------------------------------------------------
\subsection{The pulsar-SMBH timing model} \label{subsec:timing model}
%----------------------------------------------------------------------

With the pulsar's orbital motion and light propagation in the SMBH spacetime
described before, we can build the timing model that predicts the TOAs of pulsar
pulses. The pulsar rotates extremely stably in its own inertial frame.
Therefore, the rotation of the pulsar can be described by~\citep{Damour:1986}
\begin{equation}\label{eq:proper rotation}
	N(T)=N_0+\nu T+\frac{1}{2}\dot{\nu}T^2+\cdots\,,
\end{equation}
where $T$ is the proper time of the pulsar, $N$ is the proper rotation number,
$N_0$ is a constant for initial condition, $\nu$ is the rotation frequency of
the pulsar and $\dot{\nu}$ describes the spin-down of the pulsar.  Note that,
only here, the dot in fact denotes a derivative with respect to $T$.  In our
model, we ignore the glitches of the pulsar, as large glitches can be recovered
and dealt with in later data analysis, while small glitches will be likely
counted as timing red noise discussed in the later section.

We may regard the emission time of the pulsar pulse to be the time that the
pulsar rotated to a specific direction and its emission beam is pointing to the
Earth. Without losing generality, we can choose it to be the time that $N$ is an
integer~\citep{Damour:1986}, then Equation~(\ref{eq:proper rotation}) gives the
emission time of the $N$-th pulse. However, the orbital motion of the pulsar
leads to an aberration effect that in fact changes the exact rotation phase of
the pulsar when the pulsar emission beam points to the Earth, and this can be
accounted for by the so-called aberration delay~\citep{Damour:1986} 
\begin{equation}
	\Delta_{\rm A1}=-\frac{1}{2\pi\nu}\frac{\bm{v}\cdot(\hat{\bm{K}}_0\times
	\hat{\bm{e}}_3)}{c|\hat{\bm{K}}_0\times\hat{\bm{e}}_3|^2}\,,
\end{equation}
where a term proportional to $r/d$ is ignored as the distance between Sgr~A* and
SSB is large enough. For a pulsar with its spin aligned with the orbital angular
momentum, the amplitude of the leading-order aberration delay is at the order of
\begin{equation}
	\Delta_{\rm A1}\sim 3\,{\rm ms}\left(\frac{\nu}{1\,{\rm Hz}}\right)^{-1}
	\left(\frac{P_b}{1\,{\rm yr}}\right)^{-1/3}\sin^{-1}i\,,
\end{equation}
which suggests that the aberration delay should be taken into account for a
pulsar-SMBH system, especially if a normal pulsar is observed. In fact, the
next-order aberration delay only becomes smaller by a factor of
$v/c\sim10^{-1}-10^{-2}$ and thus might be relevant for a normal pulsar or a
pulsar in a tight orbit. In Appendix~\ref{app:second order aberration}, we give
a calculation of the next-order aberration delay caused by the pulsar's motion,
and it is expressed as
\begin{equation}
	\Delta_{\rm A2}=\Delta_{\rm A1}\left(\frac{\bm{v}\cdot\hat{\bm{K}}_0}{2c}
	+\frac{\bm{v}\cdot\left[\hat{\bm{K}}_0-(\hat{\bm{K}}_0\cdot\hat{\bm{e}}_3)
	\hat{\bm{e}}_3\right]}{c|\hat{\bm{K}}_0\times\hat{\bm{e}}_3|^2}
	\right)\,.
\end{equation}
Though the timing precision for normal pulsars is, in general, worse than that
for millisecond pulsars, we add this term to our timing model as it can be
relevant for pulsars in relativistic and eccentric orbits.

We remind here that the aberration, as well as the lensing of the light to be
discussed later, will also effectively change the line of sight direction for
the pulsar, which means that we will observe a different part of the emission
region as a function of time~\citep{Damour:1991rd}. This latitudinal aberration
effect has been observed in, for example, PSR~B1534$+$12~\citep{Stairs:2004ye}.
In the pulsar-SMBH system, the line of sight variation at the leading order
is~\citep{Damour:1991rd}
\begin{equation}
	\delta\zeta=-\left(\hat{\bm{K}}_0 \times \frac{\bm{v}}{c}\right) \cdot\frac{
	\hat{\bm{e}}_3 \times\hat{\bm{K}}_0}{|\hat{\bm{e}}_3 \times
	\hat{\bm{K}}_0|}\,,
\end{equation}
which is about $1^\circ$ for a pulsar with $P_b = 1\,{\rm yr}$. Depending on the
real situation, the variation of the pulse profile may or may not need to be
taken into account for a better determination of TOAs. Here, we ignore the
latitudinal effect as it relates to the modeling of the pulsar emission region
that varies for different pulsars.

The above aberration delay fully comes from special-relativistic effects in the
flat spacetime. The deflection of the pulsar's signal caused by the SMBH leads
to a lensing correction to the aberration delay, \ZX{}{which is called longitudinal deflection delay~\citep{Doroshenko:1995rm, Hu:2022jiq}}
\begin{equation}
	\Delta_{\rm L}=\frac{1}{2\pi\nu}\frac{2GM}{c^2}\frac{\hat{\bm{n}}}{r-z}\cdot
	\frac{\hat{\bm{K}}_0\times\hat{\bm{e}}_3}{|\hat{\bm{K}}_0\times
	\hat{\bm{e}}_3|^2}\,,
\end{equation}
which is proportional to the mass of SMBH. For a pulsar that spins aligned with
the orbital angular momentum, the term is at the order of $\sim 0.3\,{\rm ms}$
for a pulsar with  $\nu=1\,{\rm Hz}$, $P_b=1\,{\rm yr}$, and a moderate orbital
inclination.  Although the mass of SMBH is very large compared to the usual
companion in a binary pulsar system, the lensing correction still only becomes
important for systems close to edge-on. Nevertheless, we include this correction
in our timing model.

Counting into the aberration delay, the emission time $T_e$ of a pulse in the
pulsar's proper frame is given by 
\begin{equation}\label{eq:T2Te}
	T_e=T+\Delta_{\rm A1}+\Delta_{\rm A2}+\Delta_{\rm L}\,.
\end{equation} 
We note that the quantities in the aberration delay, such as $\bm{r}$ and
$\bm{v}$, in principle should take their values at time $T_e$, which is the time
when the pulse is emitted. However, changing to their values at $T$ only
introduces a fractional change at the order of $\Delta_{\rm A}/P_b$, which can
be safely ignored.

The proper time $T$ of the pulsar is related to the coordinate time $t$ via
\begin{eqnarray}\label{eq:dTdt}
	\frac{{\rm d}T}{{\rm d}t}& \simeq 1&-\frac{GM}{rc^2}-\frac{1}{2}\frac{v^2}{c^2}+
	\frac{1}{2}\left(\frac{GM}{rc^2}\right)^2\nonumber\\
	&&-\frac{3}{2}\frac{GM}{rc^2}\frac{v^2}{c^2}-\frac{1}{8}\frac{v^4}{c^4}\,,
\end{eqnarray}
accurate to the 2PN order~\citep{Wex:1995}. The spin contribution of the SMBH, 
\begin{equation}
	\left(\frac{{\rm d}T}{{\rm d}t}\right)_{\rm S}=\chi\frac{2G^2M^2}{c^5r^2}
	\hat{\bm{s}}\cdot(\hat{\bm{n}}\times\bm{v})\,,
\end{equation}
is ignored to be consistent with our 2PN equations of motion~(\ref{eq:equation
of motion}). Given the initial condition of $T$ at the reference time,
Equation~(\ref{eq:dTdt}) gives us the pulse emission time $t_e$ at the
center-of-mass frame. However, in pulsar timing, it is used to redefine the
pulsar's proper time and to absorb the linear dependence of $t$ in $T$, which
corresponds to a redefinition of the pulsar spin and its time derivative in the
pulsar's proper frame~\citep{Blandford:1976}. Though the effects of the SMBH spin and
quadrupole moment in the pulsar's orbital motion make it hard to strictly define
a constant scaling factor for $T$ that absorbs all the linear dependence, as
there are several different precession time scales in the system, we can still
use a leading-order approximation,
\begin{equation}
	T\rightarrow f_{\rm E}T\equiv\left(1-\frac{2GM}{ac^2}-\frac{E}{c^2}\right)T\,,
\end{equation} 
where $E$ is the orbital energy at the Newtonian order. These transformation
fully degenerates with the redefinition of $\nu$ and $\dot{\nu}$, and thus will
not affect the parameter estimation for parameters that we are interested in.
After this redefinition, one can introduce the so-called Einstein delay that
accounts for the rest difference between $T$ and $t$~\citep{Blandford:1976}, 
\begin{equation}
	\Delta_{\rm E}=t-T\,,
\end{equation}
and we calculate it by integrating 
\begin{eqnarray}\label{eq:dDeltaEdt}
	\frac{{\rm d}\Delta_{\rm E}}{{\rm d}t}=1-\frac{1}{f_{\rm E}} && \left[
	1-\frac{GM}{rc^2}-\frac{1}{2}\frac{v^2}{c^2}+
	\frac{1}{2}\left(\frac{GM}{rc^2}\right)^2\right.\nonumber\\
	&&\left.-\frac{3}{2}\frac{GM}{rc^2}\frac{v^2}{c^2}-\frac{1}{8}\frac{v^4}{c^4}
	\right]\,,
\end{eqnarray}
together with the equations of motion~(\ref{eq:equation of motion}).

The emission time of the pulse in the global coordinate can then be expressed as
\begin{equation}\label{eq:Te2te}
	t_e=T_e+\Delta_{\rm E}\,,
\end{equation}
and taking into account the propagation time delay, the TOA of pulse is 
\begin{equation}\label{eq:te2toa}
	t^{\rm TOA}=t_e+\Delta_{\rm R}+\Delta_{\rm S1}+\Delta_{\rm S2}
	+\Delta_{\rm FD}\,.
\end{equation} 
Equations~(\ref{eq:proper rotation}), (\ref{eq:T2Te}), (\ref{eq:Te2te}) and
(\ref{eq:te2toa})---together with the equations of motion for the
pulsar---compose the basic timing model for the pulsar-SMBH system. 

Before we proceed, we discuss here the effects of the proper motion of Sgr~A*.
In previous discussions, we always assume that the distance between Sgr~A* and
SSB tends to infinity. However, the distance between Sgr~A* and SSB is about
$d=8.2\,{\rm kpc}$~\citep{GRAVITY:2020gka} and the proper motion of Sgr~A* has
been precisely measured by the Very Long Baseline Array (VLBA)
observation~\citep{Reid:2020}. The apparent motion of Sgr~A* is about
$-6.4\,{\rm mas\,yr^{-1}}$ along the Galactic plane and $-0.22\,{\rm
mas\,yr^{-1}}$ toward the North Galactic Pole, which can be mainly attributed to
the Sun's orbital motion around the GC. This relative motion between Sgr~A* and
SSB causes several apparent effects in the timing observation which are
discussed below. 

The proper motion of the pulsar-SMBH system changes its geometrical orientation
with respect to the Earth, including the secular changes in the orbital
inclination angle $i$ and the longitude of orbital periastron
$\omega$~\citep{Kopeikin:1996},
\begin{subequations}
	\begin{eqnarray}
		\delta i&=& \big(-\mu_\alpha\sin\Omega + \mu_\delta\cos\Omega
		\big)t\,,\\
		\delta\omega&=&\sin^{-1}i \big(\mu_\alpha\cos\Omega+\mu_\delta\sin\Omega
		\big)t\,,
	\end{eqnarray}	
\end{subequations}
where $\mu_\alpha$ and $\mu_\delta$ are the components of the proper motion
along  $\hat{\bm{I}}_0$ and $\hat{\bm{J}}_0$ directions respectively. Note that
the effects of proper motion depend on the longitude of the ascending node
$\Omega$, which is usually not an observable in the timing observation of
systems with negligible proper motion. The proper motion breaks the rotation
symmetry of the system in the timing observation and provides the possibility of
measuring $\Omega$. Taking the proper motion of Sgr~A*, one can estimate that
the apparent change rates of $i$ and $\omega$ are at the order of $10^{-6}\,{\rm
deg\ yr^{-1}}$, which corresponds to a cumulated time delay of $\sim 10\,{\rm
ms}$ for a pulsar in a 1-yr orbit and a total observation time span of 5 years.
This suggests that in future observations, one has the possibility to measure
the proper motion effects. In our timing model, we incorporate the proper motion
by applying a coordinate rotation before we calculate the R\"{o}mer delay
$\Delta_{\rm R}$ introduced in Section~\ref{subsec:light propagation}, i.e., we
replace $\hat{\bm{K}}_0$ with
\begin{eqnarray}
	\hat{\bm{K}}_0&\rightarrow&\hat{\bm{K}}_0\cos\mu t+\left(\frac{\mu_\alpha}
	{\mu}\hat{\bm{I}}_0+\frac{\mu_\delta}{\mu}\hat{\bm{J}}_0\right)\sin\mu t\,,
\end{eqnarray}
in Equation~(\ref{eq:Romer delay}), where
$\mu=\sqrt{\mu_\alpha^2+\mu_\delta^2}$. In principle, this rotation should be
applied to the calculations of all other delays. However, the PN--proper motion
crossing terms are clearly beneath the observation precision and thus they can
be ignored.

Another effect from the finite distance between Sgr~A* and SSB is the parallaxes caused by the orbital motion of the pulsar and the Earth~\citep{Kopeikin:1995}. They are also corrections to the R\"{o}mer delay (including the Solar system part) and can be considered directly by replacing the  R\"{o}mer delay $\Delta_{\rm R}$ as
\begin{equation}
	\Delta_{\rm R}\rightarrow \Delta_{\rm R}+\Delta_{\pi P}+\Delta_{\pi M}+\Delta_{\pi \odot}\,,
\end{equation}
with
\begin{eqnarray}
	\Delta_{\pi P}&=&\frac{1}{2cd}\left(\hat{\bm{K}}_0\times \bm{r}\right)^2\,,\\
	\Delta_{\pi \odot}&=&\frac{1}{2cd}\left(\hat{\bm{K}}_0\times \bm{r}_{\rm E}\right)^2\,,\\
	\Delta_{\pi M}&=&-\frac{1}{cd}\left(\hat{\bm{K}}_0\times \bm{r}\right)\cdot\left(\hat{\bm{K}}_0\times \bm{r}_{\rm E}\right)\,,
\end{eqnarray}
where $r_{\rm E}$ is the vector pointing from the Sun to the Earth. Among the three parallactic terms, $\Delta_{\pi\odot}$ with an amplitude $\sim (1\,{\rm AU})^2/2cd\sim 0.15\,{\rm \mu s}$ is only related to the Solar system motion and is not taken into account in our model. For the other two terms, their amplitudes can be estimated as
\begin{eqnarray}
	\mbox{amplitude of } \Delta_{\pi P} &\sim&\frac{a^2}{2cd}\sim 4\,{\rm ms}\left(\frac{P_b}{1\,{\rm yr}}\right)^{4/3}\,,\\
	\mbox{amplitude of } \Delta_{\pi M} &\sim&\frac{1\,{\rm AU}}{2cd}a\sim 24\,{\rm \mu s}\left(\frac{P_b}{1\,{\rm yr}}\right)^{2/3}\,.
\end{eqnarray}
For timing precision at the millisecond level, only the term $\Delta_{\pi P}$ is relevant, which is larger for a pulsar with a longer orbital period. We add this term to our timing model, which introduces an additional parameter $d$ to the timing model. Note that, as discussed by~\citet{Kopeikin:1995}, the annual-orbital mixing term  $\Delta_{\pi M}$ can help measure $\Omega$ by breaking the rotation symmetry. However, the orbital parallax term $\Delta_{\pi P}$, which is important in the pulsar-SMBH system, preserves this symmetry.

The relative motion between Sgr~A* and SSB may also introduce apparent \ZX{}{changes} in 
the pulsar orbital period via 
\begin{equation}
	\left(\frac{\dot{P}_b}{P_b}\right)^{\rm obs}=\left(\frac{\dot{P}_b}{P_b}\right)^{\rm int}-\frac{\dot{D}}{D}\,,
\end{equation}
where \ZX{}{$D\approx 1-\hat{\bm{K}}_0\cdot (\bm{v}_{\rm SMBH}-\bm{v}_\odot)\approx 
1-\hat{\bm{K}}_0\cdot\bm{v}_\odot$ (with $\bm{v}_{\rm SMBH}$ and $\bm{v}_\odot$ respectively the velocities of the SMBH and the Solar system)} is the Doppler shift of the system~\citep{Damour:1991rd}. \ZX{}{The additional term $-\dot{D}/D$ is used to be separated into the contribution of the Galactic  gravitational acceleration and the so-called Shklovskii effect}
\begin{equation}
	-\frac{\dot{D}}{D}=\left(\frac{\dot{P}_b}{P_b}\right)^{\rm Gal}+\left(\frac{\dot{P}_b}{P_b}\right)^{\rm Shk}\,,
\end{equation}
with
\begin{eqnarray}
	\left(\frac{\dot{P}_b}{P_b}\right)^{\rm Gal}&=&-\frac{1}{c}\hat{\bm{K}}_0\cdot  \bm{a}_\odot\,,\\
	\left(\frac{\dot{P}_b}{P_b}\right)^{\rm Shk}&=&\frac{\bm{v}_\odot^2-(\hat{\bm{K}}_0\cdot\bm{v}_\odot)^2}{cd}\,,
\end{eqnarray}
where we have dropped the SMBH terms.
However, considering that the motion of the Sun 
is largely described by a circular motion around the GC, the Shklovskii 
effect~\citep{Shklovskii:1970} and the Galactic gravitational acceleration effect will 
be canceled \ZX{}{due to the fact that $\hat{\bm{K}}_0\cdot\bm{v}_\odot\approx 0$ and 
$\bm{v}_\odot^2/d\approx \hat{\bm{K}}_0\cdot  \bm{a}_\odot$} and thus can be ignored in the timing model. 
\ZX{}{In fact, from the definition of $D$ one can easily see that for a circular motion $\dot{D}=0$ and there is no effect.}

%----------------------------------------------------------------------
\begin{figure}[ht!]
	\begin{center}
		\includegraphics[width=8.6cm]{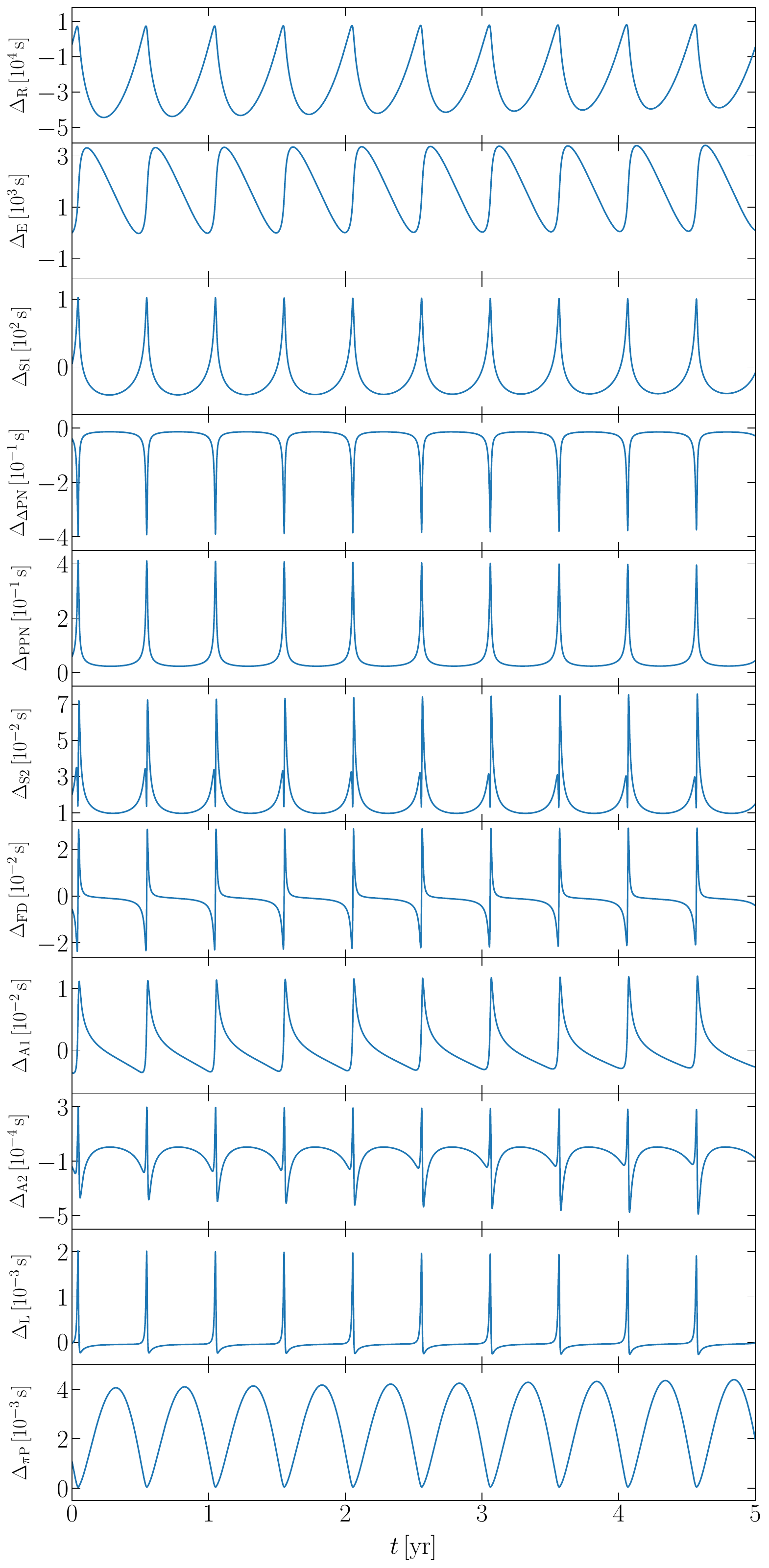}
	\caption{Various time delays described in the text for the fiducial
	pulsar-SMBH system. A constant term is removed from the 1PN Shapiro delay
	with respect to Equation~(\ref{eq:1PN Shapiro}).  \label{fig:time delay}}
	\end{center}
\end{figure}
%----------------------------------------------------------------------

As an illustration, in Figure~\ref{fig:time delay} we plot the various time
delays in the timing model for the system with parameters in
Equation~(\ref{eq:system params}). To calculate the aberration delay and \ZX{}{finite distance} effects introduced in this subsection, we have further set the parameters
of the pulsar, $\Theta_{\rm
PSR}=\left\{N_0,\nu,\dot{\nu},\lambda_{p},\eta_p\right\}$, to be
\begin{eqnarray}\label{eq:pulsar params}
	N_0&=&0.5\,, \quad \nu=1\,{\rm Hz}\,, \quad \dot{\nu}=10^{-15}\,{\rm s^{-2}}\,,\nonumber\\
	\lambda_p&=&\pi/5\,, \quad \eta_p=0\,,
\end{eqnarray}
where $\lambda_p$ and $\eta_p$ give the spin direction of the pulsar, and we set
it to be  in parallel with the orbital angular momentum in the fiducial case. We
also add the proper motion $\mu_\alpha$ and $\mu_\delta$\ZX{}{as well as the distance $d$} to the SMBH parameter
set $\Theta_{\rm SMBH}$. \ZX{}{We use the values} in~\citet{Reid:2020},
\begin{equation}\label{eq:proper motion params}
	\mu_\alpha=-3.2\,{\rm mas\,yr^{-1}}\,,\mu_\delta=-5.6\,{\rm mas\,yr^{-1}}\,,
\end{equation}
\ZX{}{and~\citet{Gravity:2019nxk}}
\begin{equation}\label{eq:distace params}
	d=8.18\,{\rm kpc}\,.
\end{equation}
%%

%----------------------------------------------------------------------
\subsection{The inverse timing model} 
\label{subsec:inverse timing model}
%----------------------------------------------------------------------

In previous subsections, we have built a timing model for pulsar-SMBH systems
accurate to the 2PN level. The timing model predicts the TOA of a pulse emitted
at a given pulsar proper time, for example, the time when pulsar's rotation
number $N$ is an integer. By fitting the predictions to the observed TOAs, one
can obtain the best-fit parameters in the timing model. However, in the
real pulsar timing process, the direct observable is the TOAs of the pulses. Due to the fact that telescopes
will not observe a pulsar continuously to record every single pulse, we, in
fact, do not know the pulsar rotation number $N$ corresponding to the observed
TOAs. Thus, the timing residual, which represents the difference between the
model prediction and observation, is defined to be~\citep{Damour:1986}
\begin{equation}\label{eq:residual}
	\mathcal{R}=\frac{1}{\nu} \Big[\mathcal{N} \big(t^{\rm TOA};\{\Theta\}\big)-
	\big\lfloor \mathcal{N} \big(t^{\rm TOA};\{\Theta\}\big) \big\rceil \Big]\,,
\end{equation}
where $\mathcal{N} \big(t^{\rm TOA};\{\Theta\} \big)$ is the predicted pulsar's
rotation number for the pulse arrived at $t^{\rm TOA}$ and with model parameters
$\{\Theta\}$; note that it needs not to be an integer \ZX{}{as the timing model may not be perfect, the model parameters may deviate from the true parameters, and  the $t^{\rm TOA}$ can contain noise}. $\big\lfloor
\mathcal{N}\big(t^{\rm TOA};\{\Theta\}\big) \big\rceil$ is the integer closest
to $\mathcal{N}\big(t^{\rm TOA};\{\Theta\}\big)$. This definition folds the
timing residual into one pulsar's rotation period, and thus it is necessary to
have a timing model accurate than the pulsar's rotation period.
Equation~(\ref{eq:residual}) requires the calculation of pulsar rotation number
from a given TOA, and the function $\mathcal{N} \big(t^{\rm TOA};\{\Theta\}
\big)$ is realized by the so-called inverse timing model~\citep{Damour:1986}.
For analytical timing models that adopt a PN expansion, the inverse timing model
can be obtained by solving Equations~(\ref{eq:T2Te}), (\ref{eq:Te2te}), and
(\ref{eq:te2toa}) with iteration~\citep{Damour:1986}. However, to invert the
numerical timing model, in general, one may need to calculate a series of TOAs
corresponding to different $N$ and inverse the numerical relation by
interpolation~\citep{DellaMonica:2025ent}. The choice of the sample rate of the
numerical relation and the interpolation procedure may introduce additional
numerical errors to the inverse timing model. Besides, a higher sample rate
generally needs more computational resources, which may not be always available
in parameter estimation with, say, Monte Carlo techniques, which needs many
calculations of the inverse timing model with different model parameters. 

\allowdisplaybreaks

\citet{Hu:2023ubk} proposed a method for building the inverse timing model
efficiently. Instead of integrating Equation~(\ref{eq:equation of motion}), one
can transform the integration variable from coordinate time $t$ to $t^{\rm TOA}$
by taking the advantage that Equation~(\ref{eq:te2toa}) has an analytical
expression that only depends on the pulsar's position at time $t$.\footnote{Here
we omit the lower index $e$ and take this equation as a general relation between
coordinate time $t$ and arrival time $t^{\rm TOA}$.} Writing out explicitly, one
has
\begin{eqnarray}\label{eq:dTOAdt}
	\frac{{\rm d}t^{\rm TOA}}{{\rm d}t}& = &1+\frac{{\rm d}\Delta_{\rm R}}{{\rm
	d}t} +\frac{{\rm d}\Delta_{\rm S}}{{\rm d}t} +\frac{{\rm d}\Delta_{\rm
	FD}}{{\rm d}t} +\frac{{\rm d}\Delta_{\pi P}}{{\rm d}t}\,,
\end{eqnarray}
where
\begin{subequations}
	\begin{eqnarray}
		\frac{{\rm d}\Delta_{\rm R}}{{\rm d}t}&=&\frac{1}{c} \left[v_z\cos\mu
		t+\left( \frac{\mu_\alpha}{\mu} v_x+\frac{\mu_\delta}{\mu}
		v_y\right)\sin\mu t\right] \nonumber\\
		&&+\frac{1}{c}\Big[\big(\mu_\alpha x+\mu_\delta y\big) \cos \mu t -\mu
		z\sin\mu t \Big]\,,\\
		\frac{{\rm d}\Delta_{\rm S}}{{\rm d}t}&=& \frac{{\rm d}\Delta_{\rm
		S1}}{{\rm d}t}+\frac{{\rm d}\Delta_{\rm \Delta PN}} {{\rm
		d}t}+\frac{{\rm d}\Delta_{\rm  PPN}}{{\rm d}t}\,,\\
		\frac{{\rm d}\Delta_{\rm S1}}{{\rm
		d}t}&=&-\frac{2GM}{c^3}\frac{\dot{r}-v_z} {r-z}\,,\\
		\frac{{\rm d}\Delta_{\rm \Delta PN}}{{\rm d}t}&=&\frac{4G^2M^2}{c^5}
		\frac{\dot{r}-v_z}{(r-z)^2}\,,\\
		\frac{{\rm d}\Delta_{\rm  PPN}}{{\rm d}t}&=&\frac{G^2M^2}{4c^5}\left[
		-\frac{v_z}{r^2}+\frac{2z\dot{r}}{r^3}+15\frac{v_z-\dot{r}z/r}{r^2-z^2}
		\right.\nonumber\\
		&&-\left. 15\arccos\left(-\frac{z}{r}\right)\frac{r\dot{r}-zv_z}
		{(r^2-z^2)^{3/2}}\right]\,,\\
		\frac{{\rm d}\Delta_{\rm FD}}{{\rm d}t} &=& -\chi
		\frac{2G^2M^2}{c^5}\left\{ \frac{\hat{\bm{s}}\cdot \left[\hat{\bm{K}}_0
		\times(\bm{v}/r-\hat{\bm{n}} \dot{r}/r)\right]}{r-z}\right. \nonumber\\
		&& -\left.\frac{\hat{\bm{s}}\cdot\left(\hat{\bm{K}}_0\times\hat{\bm{n}}
		\right)}{r-z}\frac{\dot{r}-v_z}{r-z}\right\}\,,\\
		\frac{{\rm d}\Delta_{\pi P}}{{\rm d}t} & = &
		\frac{1}{cd}(xv_x+yv_y)\,, 
	\end{eqnarray}	
\end{subequations}
with $(x,y,z)$ and $(v_x,v_y,v_z)$ the coordinate components of pulsar position
and velocity, respectively. In the R\"{o}mer delay, we have taken into account
the proper motion of Sgr~A* as discussed before. 

Using Equation~(\ref{eq:dTOAdt}), one may integrate the following system 
\begin{eqnarray}\label{eq:invMotion}
	\frac{{\rm d}}{{\rm d}t^{\rm TOA}}
	\begin{pmatrix}
		\bm{r}\\
		\bm{v}\\
		t\\
		\Delta_{\rm E}
	\end{pmatrix}
	&=&
	\begin{pmatrix}
		\bm{v}\\
		\ddot{\bm{r}}\\
		1\\
		{\rm d}\Delta_{\rm E}/{\rm d}t
	\end{pmatrix}
	\frac{{\rm d}t}{{\rm d}t^{\rm TOA}}(\bm{r},\bm{v},t)\,,
\end{eqnarray}
with $\ddot{r}$ and ${\rm d}\Delta_{\rm E}/{\rm d}t$ given by
Equations~(\ref{eq:equation of motion}) and (\ref{eq:dDeltaEdt}), respectively.
The initial condition for the above differential system can also be transformed
from the original initial condition. Here, we take the reference time to be
$t^{\rm TOA}=0$ when the pulsar has initial position $\bm{r}_0$ and velocity
$\bm{v}_0$ given by its orbital elements. One can calculate the corresponding
coordinate time $t$ with the equations of various time delays. Finally, we
choose $\Delta_E=0$ at the reference time.

Integrating the system~(\ref{eq:invMotion}) then gives the system directly as
functions of $t^{\rm TOA}$. Given a TOA, one can read out the associated pulse
emission time $t$ and Einstein delay $\Delta_{\rm E}$ directly from the
integration output. One can then calculate $T_e$, $T$, and $N$ using
Equations~(\ref{eq:Te2te}), (\ref{eq:T2Te}), and (\ref{eq:proper rotation})
sequentially. This method fully utilize the advantage of the adaptive
integration method, as in general the TOA observation cadence is much longer
than the integration time step that is needed for the desired integration
precision. 

In Appendix~\ref{app:numerical accuracy of the timing model}, we give a more
detailed discussion of the numerical accuracy of our code.

%----------------------------------------------------------------------
\section{Parameter estimation} \label{sec:parameter estimation}
%----------------------------------------------------------------------

Based on the realistic timing model constructed in the previous section, we are
able to forecast the future timing observation of possible pulsar-SMBH systems.
In this section, we discuss the measurability of various system parameters using
the Fisher matrix method. Some of these results were obtained in previous
studies~\citep{Liu:2011ae,Psaltis:2015uza,Hu:2023ubk} based on timing models
that only take into account some of the previously discussed effects. Those
earlier results remain valid when higher-order terms do not \ZX{}{affect} the parameter
estimation dramatically. Nevertheless, we still present more realistic results
here as a verification and for completeness. Further, we discuss the effects
related to the aberration and proper motion, which are considered for the first
time in pulsar-SMBH systems. This section is constructed as follows. In
Section~\ref{subsec:Fisher setup}, we introduce the parameter estimation method
and numerical setup for our analysis. In Section~\ref{subsec:MSQ estimation}, we
forecast the measurability of the parameters of the GC SMBH, Sgr~A*. We discuss
the effect of the proper motion of Sgr~A* in Section~\ref{subsec:proper motion}.
The detectability of the pulsar's spin direction based on aberration effects is
discussed in Section~\ref{subsec:aberration}.

%----------------------------------------------------------------------
\subsection{The Fisher matrix method}\label{subsec:Fisher setup}
%----------------------------------------------------------------------

As we introduced before, the timing residual~(\ref{eq:residual}) for a pulse
arrived at $t^{\rm TOA}$ can be calculated with the inverse timing model after
giving all the system parameters $\Theta$,
\begin{equation}
	\Theta=\Theta_{\rm SMBH}\cup\Theta_{\rm orb}\cup\Theta_{\rm PSR}\,,
\end{equation}
where
\begin{subequations}\label{eq:system parameters}
\begin{eqnarray}
	\Theta_{\rm SMBH}&=& \big\{M, \, \chi, \, \lambda, \, \eta, \, q, \,
	\mu_\alpha, \, \mu_\beta \,,d\big\}\,,\\
	\Theta_{\rm orb}&=& \big\{P_b, \, e, \, i, \, \omega, \, \Omega, \, T_0
	\big\}\,,\\
	\Theta_{\rm PSR}&=& \big\{N_0, \, \nu, \, \dot{\nu}, \, \lambda_p, \, \eta_p
	\big\}\,.
\end{eqnarray}
\end{subequations}
However, for separating out the rotation symmetry, we in fact use $\eta-\Omega$
and $\eta_p-\Omega$ instead of $\eta$ and $\eta_p$. As we will discuss later,
even with known proper motion, the measurement precision of $\Omega$ is not
high. 

Assuming a Gaussian timing noise in the observation, the probability for the
system to have parameters $\Theta$ is
\begin{equation}\label{eq:probability}
	P \big(\Theta|\{t^{\rm TOA}\} \big)
	\propto\exp\left(-\frac{1}{2}\sum_{i=1}^{N_{\rm TOA}} \frac{\mathcal{R}
	\big(t^{\rm TOA}_i;\Theta \big)^2}{\sigma^2_{{\rm TOA},i}}\right)\,,
\end{equation}
where the expression for $\mathcal{R}$ is given in Equation~(\ref{eq:residual})
and the summation is over all observed TOAs denoted by $\{t^{\rm TOA}\}$.
$\sigma_{{\rm TOA},i}$ is the timing precision for the $i$-th TOA. As discussed
before, due to the fact that we cannot know the true value of the pulsar's
rotation number related to an observed pulse, in the right-hand side of
Equation~(\ref{eq:residual}), the second term is chosen to be the closest
integer to the first term. However, in the Fisher matrix analysis of simulated
data sets, it will be convenient to use $N \big(t^{\rm TOA};\bar{\Theta} \big)$
instead of the second term, where $\bar{\Theta}$ represents the true value of
the system parameters that are used in simulations. This replacement will be
valid if the timing noise and model inaccuracy are smaller than the pulsar's
rotation period.

As a standard procedure, the covariance matrix in the Fisher formalism, defined
by
\begin{equation}\label{eq:cov matrix1}
	C_{\alpha\beta}=\left(\frac{\partial^2 \mathcal{L}}{\partial\Theta^\alpha
	\partial\Theta^\beta}\right)^{-1}\,,
\end{equation}
can serve to obtain an estimation of the measurement uncertainties of the
parameters, where $\mathcal{L}=-\ln P \big(\Theta \big|\{t^{\rm TOA}\} \big)$ is
the log-likelihood function. For the purpose of estimation, and considering the
replacement discussed before, Equation~(\ref{eq:cov matrix1}) can be well
approximated by
\begin{equation}\label{eq:cov matrix2}
	C_{\alpha\beta}=\left(\frac{1}{\nu^2}\sum_{i=1}^{N_{\rm TOA}}
	\frac{1}{\sigma^2_{{\rm TOA}, i}} \frac{\partial \mathcal{N}_i} {\partial
	\Theta^\alpha}\frac{\partial \mathcal{N}_i}{\partial
	\Theta^\beta}\right)^{-1}\,,
\end{equation}
where $\mathcal{N}_i=\mathcal{N}(t^{\rm TOA}_i;\Theta)$, and the above equation
is evaluated at $\bar{\Theta}$. Equation~(\ref{eq:cov matrix2}) is equivalent to
Equation~(\ref{eq:cov matrix1}) if no noise is present, and it only requires the
evaluation of the first-order derivative of the inverse timing model. We shall
note that the numerical differentiation performed here needs additional care,
and more numerical details are discussed in Appendix~\ref{app:numerical detail}.

Finally, we shall notice that, though we list all the system parameters in
Equation~(\ref{eq:system parameters}), for timing-only observations and distant
pulsars, there always exists a rotation symmetry around the line of sight. In
principle, one needs fix at least one parameter among $\Omega$, $\eta$,
$\mu_\alpha$, $\mu_\delta$, and $\eta_p$. Considering that $\eta_p$ is only
related to the aberration effect, and $\eta$ in general is considered together
with the other two spin parameters, in the later discussion, we group $\Omega$
and the proper motion parameters, and will always fix $\Omega$, or $\mu_\alpha$
and $\mu_\delta$.

%----------------------------------------------------------------------
\subsection{Measurability of SMBH parameters}
\label{subsec:MSQ estimation}
%----------------------------------------------------------------------

Now we forecast the measurability of the SMBH mass, spin, and quadrupole moment
based on mock data simulations. As we will discuss later, here we choose to
treat the  proper motion of SMBH as known parameters, i.e., we fix $\mu_\alpha$
and $\mu_\delta$ for parameter estimation here. Though we have considered
higher-order effects in our timing model, the measurement precision of SMBH
mass, spin, and quadrupole moment should not show significant differences from
previous studies that used a  timing model with only some leading-order
effects~\citep{Liu:2011ae,Psaltis:2015uza,Hu:2023ubk}. This can serve as a check
for our numerical procedure. In reality, when the effects from higher-order
terms are larger than the timing precision, our timing model should be used.

We simulate a system with parameters given in Equations~(\ref{eq:system
params}), (\ref{eq:pulsar params}), and (\ref{eq:proper motion params}), for
weekly observations spanning 5 years, which in total give 260 TOAs\ZX{}{assuming that 
we get one TOA per observation}. The timing
precision of a normal pulsar in the GC region for future SKA-like telescopes may
reach $1\,{\rm ms}$ or even better~\citep{Liu:2011ae}, thus we assume that the
timing precision $\sigma_{\rm TOA}=1\,{\rm ms}$ for all TOAs. Other precision
can be easily rescaled from our results.

%----------------------------------------------------------------------
\begin{figure*}[t]
	\begin{center}
		\includegraphics[width=17cm]{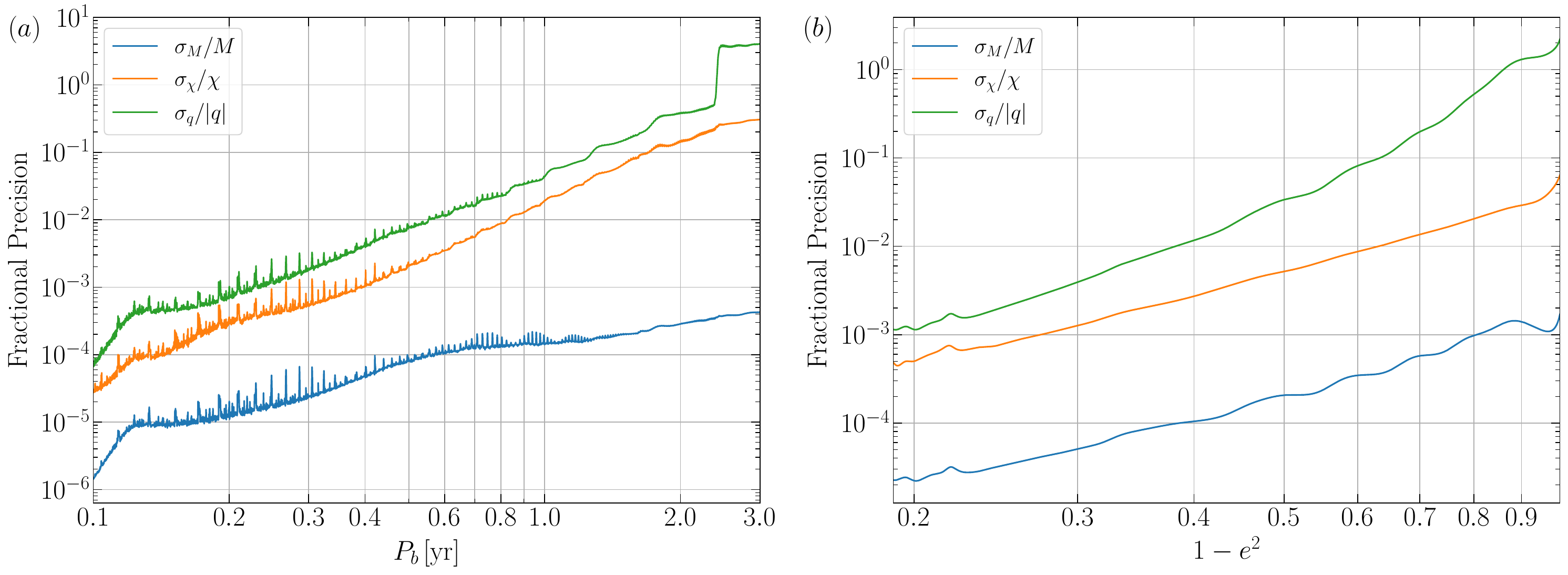}
	\caption{The fractional precision of SMBH parameters as functions of the
	pulsar's (a) orbital period and (b) orbital eccentricity.
	\label{fig:full_PE_Pb}}
	\end{center}
\end{figure*}
%----------------------------------------------------------------------

In Figure~\ref{fig:full_PE_Pb}~(a), we show the fractional precision of the SMBH
parameters as functions of the pulsar orbital period. As expected in previous
studies~\citep{Liu:2011ae,Psaltis:2015uza,Hu:2023ubk}, for pulsars with orbital
periods $P_b\lesssim 0.5\,{\rm yr}$, the measurement precision of SMBH mass,
spin, and quadrupole moment can all be better than $1\%$ level, which can
provide a stringent test of the no-hair theorem. 

One may notice that there are lots of spike structures in the curves shown in
Figure~\ref{fig:full_PE_Pb}~(a). As verified in Appendix~\ref{app:numerical detail},
we reckon that these structures are at least real results for the Fisher matrix
approximation instead of being caused by numerical errors. In this figure, for
each curve we plot $5000$ points so that one can see detailed structures. These
spikes are centered around positions where $P_b$ is an integer (rational) multiple of
weeks. As in the parameter estimation, we assumed an observational cadence of
exactly one week. In real observations, one will not expect the pulsar-SMBH
system to locate in such a peak, and the observational cadence will also be
more complex and relatively irregular, which will smooth these structures. In
contrast, the step-like structure around $P_b\sim 2.5\,{\rm yr}$ is due to an
additional periastron passage being observed, which will also occur for real
observations where the total time spans only cover one or two orbits.

Figure~\ref{fig:full_PE_Pb} (b) shows the fractional precision of the SMBH parameters
as functions of the orbital eccentricity, while the pulsar orbital period is
fixed to $P_b=0.5\,{\rm yr}$. Same as before, each curve here contains 5000
points. As we discussed in \citet{Hu:2023ubk}, the fractional precision depends
on $1-e^2$ in a nearly power-law manner. All the results shown here are
consistent with previous studies.

%----------------------------------------------------------------------
\subsection{The effects of proper motion}\label{subsec:proper motion}
%----------------------------------------------------------------------

In previous figures, we showed the results that all parameters in
Equation~(\ref{eq:system parameters}), except the proper motion parameters
$\mu_\alpha$ and $\mu_\delta$,\ZX{}{as well as the distance $d$,} are treated as free parameters. The proper motion
of Sgr~A* breaks the rotation symmetry around the line of sight and specifies a
sky orientation, enabling a measurement of the longitude of the ascending node
$\Omega$. Previous studies suggested that a measurement of $\Omega$ may help us
break the leading-order degeneracy in the measurement of spin
components~\citep{Zhang:2017qbb,Hu:2023ubk}. Therefore we study the effects of
the proper motion here. \ZX{}{As discussed before, the finite distance $d$ introduces the parallactic terms, among which the annual-orbital term could also break the rotation symmetry~\citep{Kopeikin:1995}. However, for the pulsar-SMBH system and the assumed timing precision, only the orbital term $\Delta_{\pi P}$ is relevant. Nevertheless, we also study the measurability of the distance $d$ with timing-only observation.}

%----------------------------------------------------------------------
\begin{figure*}[t]
	\begin{center}
		\includegraphics[width=17cm]{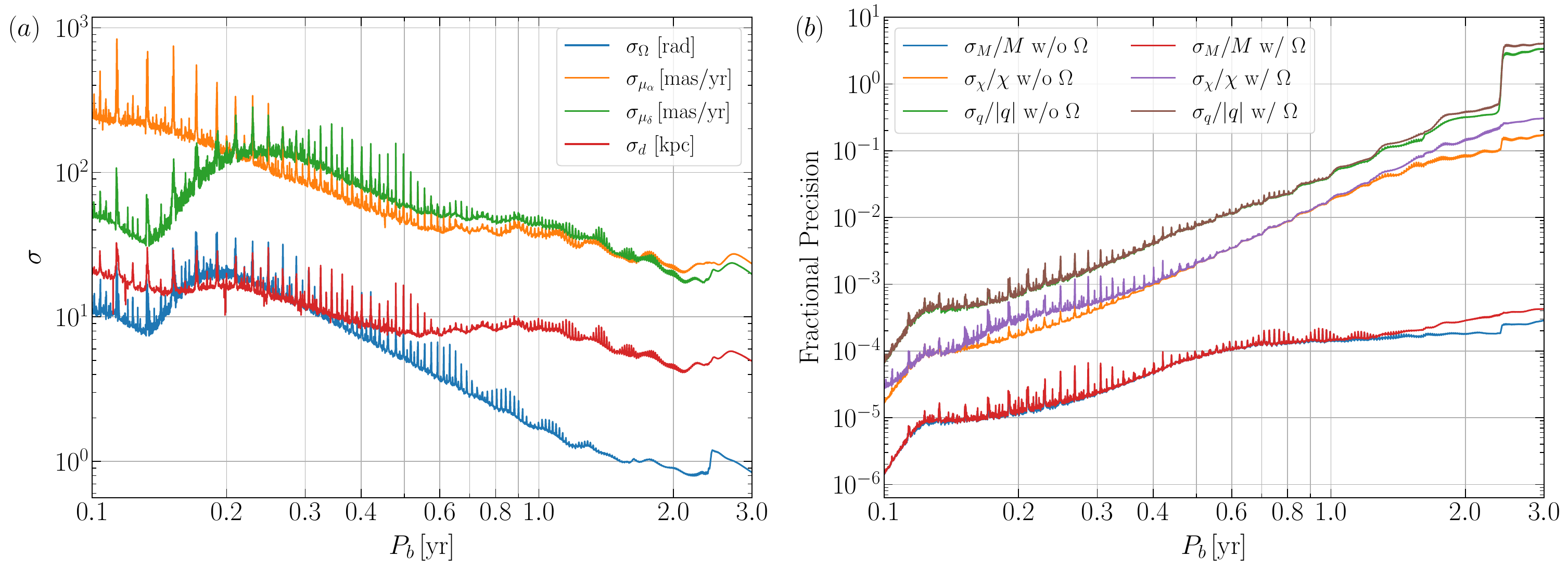}
	\caption{(a) The measurement precision of the proper motion parameters
	$\mu_\alpha$ and $\mu_\delta$, the longitude of the ascending node
	$\Omega$, \ZX{}{and the distance $d$}. Due to the degeneracy caused by the rotation symmetry, for
	estimating the measurability of the proper motion parameters, we fix
	$\Omega$, and vice versa, we fix the proper motion parameters when
	estimating the measurement precision of $\Omega$. (b) Comparison of
	parameter estimation results with or without considering the proper-motion
	effect. For parameter estimation that does not consider the proper motion,
	we fix $\Omega=0$ without losing generality.  \label{fig:full_PE-pm_Pb}}
	\end{center}
\end{figure*}
%----------------------------------------------------------------------

We first study the measurement precision of $\Omega$, or the proper motion
parameters $\mu_\alpha$ and $\mu_\delta$, when only timing data are considered.
As mentioned before, the VLBA observations have measured the proper motion of
Sgr~A* to a relatively high precision~\citep{Reid:2020}, therefore $\mu_\alpha$
and $\mu_\delta$ can be treated as fixed parameters in timing data analysis.
Nevertheless, one can also fix $\Omega$ to obtain a timing-only measurement of
the proper motion; in this case, the direction of the proper motion is
referenced to the pulsar's orbit. In principle, only when the timing-only
measurement precision is much worse than the VLBA measurement precision, the
previously adopted treatment is valid. In Figure~\ref{fig:full_PE-pm_Pb}~(a), we show the
expected measurement precision of the proper motion parameters and the longitude
of the ascending node from timing observation. \ZX{}{Similar to the proper motion, we 
show the timing-only measurement precision of the distance $d$.} Due to the degeneracy caused by
the rotation symmetry, when estimating the proper motion parameters, we fix the
longitude of the ascending node, and vice versa. One can see that the expected
measurement precisions of the proper motion parameters $\mu_\alpha$ and
$\mu_\delta$ are worse than $10\,{\rm mas/yr}$ from timing-only observation,
which is far worse than the VLBA observation. Therefore, it is better to fix the
value of the proper motion parameters to the VLBA measurement, which accounts
for the proper-motion effect with enough accuracy. \ZX{}{Similar argument applies 
to the distance $d$, which can be fixed to the measured value as the timing-only observation 
has a much weaker constraint.}

Fixing the proper motion parameters then enables a measurement of the longitude
of ascending node $\Omega$ of the pulsar's orbit, as the proper motion breaks
the rotation symmetry~\citep{Kopeikin:1996}. However, from the figure we can see
that, for pulsars with small orbits, one can still hardly determine the value of
$\Omega$.\footnote{Note that the unit of $\sigma_\Omega$ in the figure is
radian.} For pulsars with larger orbits, the measurement precision in $\Omega$
becomes better. As discussed in Section~\ref{subsec:timing model}, the effects
of the proper motion are effectively reflected in the R\"{o}mer delay, which is
proportional to the orbital length scale, while the precession rate caused by
proper motion effects is fixed.

Previous studies have pointed out that there is a leading-order degeneracy among
the spin parameters, $\{\chi,\lambda,\eta\}$, of the SMBH. This degeneracy
largely limits the measurement precision of the SMBH
spin~\citep{Liu:2011ae,Zhang:2017qbb,Hu:2023ubk, Hu:2024blq}. The leading-order
degeneracy also originates from the rotation symmetry in the pulsar timing
technique. Roughly speaking, the spin-orbit coupling causes additional
precession in $i$, $\omega$, and $\Omega$. However, for timing observation in a
short time span, only $\dot{i}$ and $\dot{\omega}$ are observable, which cannot
fully determine the three spin parameters~\citep{Liu:2011ae}. Therefore, to
break this degeneracy, one possibility argued in previous studies is to combine
the proper motion information~\citep{Zhang:2017qbb}, which enables a measurement
of $\dot{\Omega}$. However, from Figure~\ref{fig:full_PE-pm_Pb}~(a) one can see
that, even with known proper motion parameters, for pulsars with short orbital
periods, the measurement precision of $\Omega$ is quite low (e.g., for
$P_b\lesssim 1\,{\rm yr}$,  one has $\sigma_{\Omega}\gtrsim 2\,{\rm rad}$). This
low precision suggests that for pulsars with small orbital periods, introducing
the proper-motion effect cannot break the leading-order degeneracy efficiently.
For pulsars with much larger orbital periods, as studied
by~\citet{Zhang:2017qbb}, combining the astrometric observation might help. In
real observations, one may also break this leading-order degeneracy by combining
the timing observation of two or more pulsars if they are
observed~\citep{Hu:2024blq}.

In Figure~\ref{fig:full_PE-pm_Pb}~(b), we give a comparison of the parameter estimation
results with and without considering the proper-motion effects and parameter
$\Omega$. In previous studies, without considering the proper motion of Sgr~A*,
$\Omega$ is not measurable, so that it is excluded from parameter estimation.
However, for real observations, the proper motion of Sgr~A* can be
non-negligible, especially for pulsars with large orbital periods. Therefore,
including parameter $\Omega$ in parameter estimation might be necessary. From
the figure, one can see that the proper motion only introduces a small
difference in the measurement precision of SMBH parameters, as it is a small
effect. Also, as expected, the influence is generally larger for pulsars with
large orbital periods. Nevertheless, in this work, we treat $\Omega$ as a free
parameter that needs to be estimated. 

%----------------------------------------------------------------------
\subsection{The aberration effects}
\label{subsec:aberration}
%----------------------------------------------------------------------

In Section~\ref{subsec:timing model}, we discussed the aberration delays,
including $\Delta_{\rm A1}$, $\Delta_{\rm A2}$ and $\Delta_{\rm L}$, which come
from the rotation origin of the pulsar's ``lighthouse'' pulse. \ZX{}{As} discussed
in~\citet{Damour:1986}, the leading-order aberration effect will be absorbed in
a redefinition of various timing parameters, we have to include them here for
the following reasons. First, different from the DD timing model constructed
by~\citet{Damour:1986}, which parameterized the various effects in a
theory-independent way, the timing model discussed in this paper is directly
based on the physical parameters. Each of the parameters might account for
several effects both in the pulsar orbital motion and light propagation.
Therefore, the aberration effect might not be able to be fully absorbed by other
parameters in our pulsar-SMBH timing model. Further, as shown in the previous
section, the amplitude of the aberration delay can be as large as $10\,{\rm
ms}$, which is larger than the \ZX{}{assumed} timing precision. As argued
by~\citet{Damour:1986} and~\citet{Kramer:2021jcw}, the aberration delay needs to
be considered in order to give an unbiased parameter estimation result. 

However, different from other effects, all three aberration effects include the
additional parameters $\lambda_p$ and $\eta_p$ that describe the direction of
the pulsar spin axis $\hat{\bm{e}}_3$. As discussed in
Section~\ref{subsec:orbital dynamics}, the precession time scale of the pulsar
spin axis is much longer than the corresponding time scale in usual
comparable-mass binary pulsar systems.
Therefore, it might be hard to obtain the geometry of the pulsar using
pulse-profile analysis~\citep{Ferdman:2013xia}. The direction of the pulsar spin
axis then needs to be determined from the timing observation. 

%----------------------------------------------------------------------
\begin{figure}[t]
	\begin{center}
		\includegraphics[width=8.6cm]{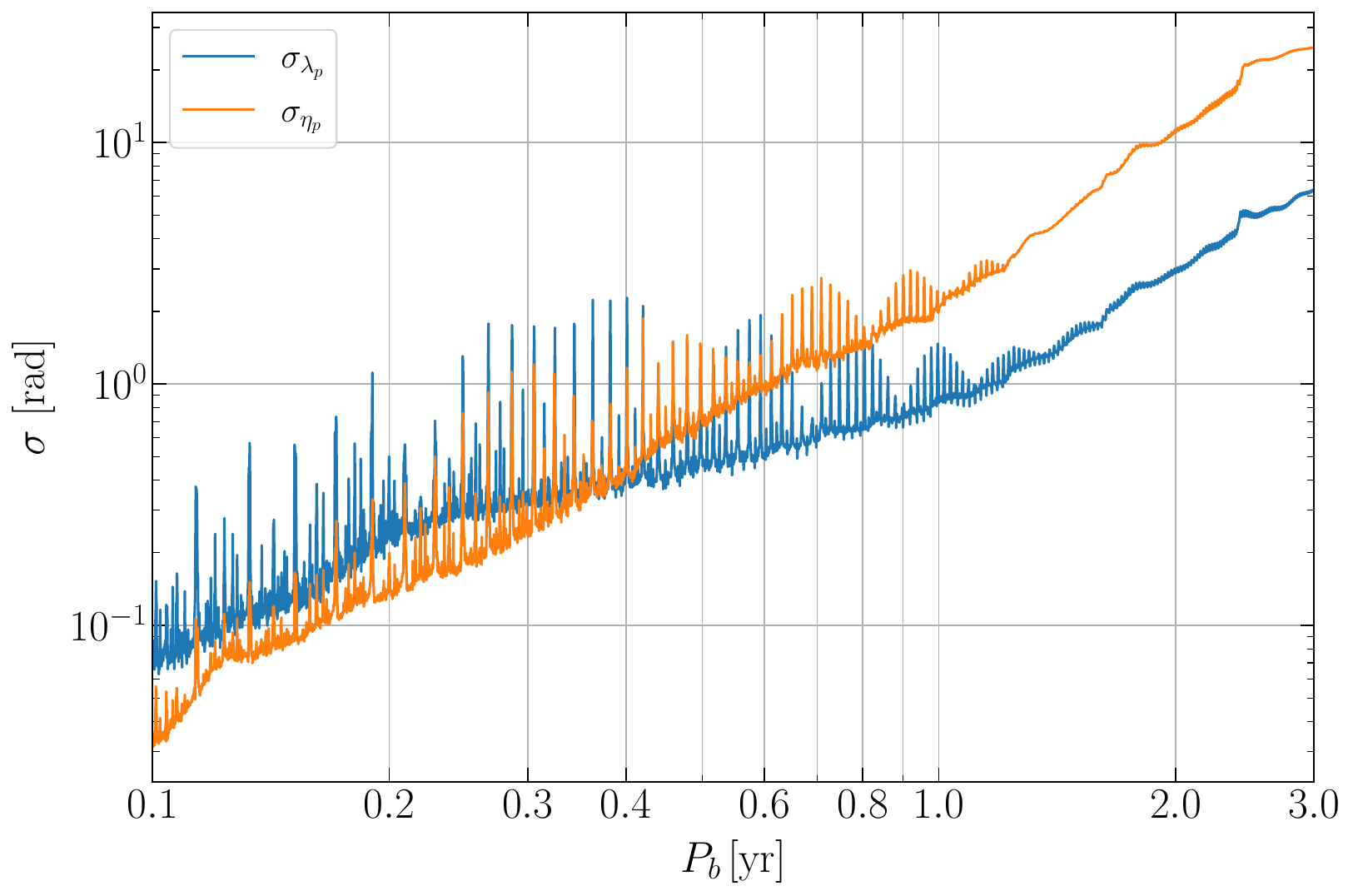}
	\caption{Expected measurement precision of $\lambda_p$ and $\eta_p$ as
	functions of the pulsar orbital period $P_b$.  \label{fig:full_PE_e3_Pb}}
	\end{center}
\end{figure}
%----------------------------------------------------------------------

In Figure~\ref{fig:full_PE_e3_Pb}, we plot the expected measurement precision of
the parameters $\lambda_p$ and $\eta_p$ as functions of the pulsar's orbital
period. As the aberration effects are suppressed by the factor of $P/P_b$, where
$P$ is the rotation period of the pulsar, for larger orbital periods, the
constraints on $\lambda_p$ and $\eta_p$ are weaker. For pulsars with orbital
periods $P_b\lesssim1\,{\rm yr}$, one might measure the directional angle of the
pulsar spin axis to a precision around or better than $1\,{\rm rad}$ through
timing-only observations for a reasonable observation time span. 

%----------------------------------------------------------------------
\section{Presence of timing red noise}\label{sec:red noise}
%----------------------------------------------------------------------

The parameter estimation in the previous sections was based on the probability
function given in Equation~(\ref{eq:probability}), where we assumed a Gaussian
noise realization. However, it is well known that for pulsar timing, there can
exist a red noise component in the timing residual due to various
reasons~\ZX{}{\citep{Hobbs:2009vh,Shannon:2012tr,Lentati:2017cbh}}. Apart from the main contribution that comes from
the rotation irregularity of the pulsar itself, for pulsars in the GC region,
the dispersion measure variation and strong scattering of the interstellar
medium may also introduce additional red noise. 

The procedures of timing analysis in the presence of red noise have long been
developed~\citep{Coles:2011zs,Lentati:2012xb,Parthasarathy:2019txt}, as it is
important for studying the pulsar physics~\citep{Greenstein:1970,Lyne:2010ad} as
well as understanding their sensitivity to nHz-frequency gravitational
waves~\citep{EPTA:2015ike,Lentati:2016ygu,Lam:2016iie,NANOGrav:2023gor}. In the
context of testing gravity theories using binary pulsars, the red noise is often
ignored, as the best results usually come from millisecond
pulsars~\citep{Kramer:2021jcw}, which have much smaller intrinsic red noise
compared to normal pulsars. Further, the tight orbit of a binary pulsar usually
required by testing relativistic gravity also makes a clear separation between
the interested time scale---in most cases, the orbital time scale---and the red
noise time scale, say, the observation time span, so that the red noise effects
on the interested parameters are negligible in many situations~\citep{Kramer:2021jcw}.

However, for a pulsar orbiting around Sgr~A* considered in this paper, it might
be worthwhile to study the red noise effects in timing analysis. As mentioned
above, for pulsars in the GC region, there might be red noise contributions from
the complex environment. Further, it is expected that we are more likely to find
normal pulsars in that region from an observational point of view, because the
interstellar scattering may prevent identifying the pulses of millisecond pulsars
\citep {Schoedel:2024}. Therefore, if we find a normal pulsar, a large intrinsic
timing red noise of the pulsar may exist. A pulsar-SMBH system is also different
from the usual binary pulsar systems in its orbital time scale, which can be on
the order of years. For a 5-yr observation time span considered in this paper,
the red noise time scale is not clearly separated from the orbital time scale,
thus it might cause larger effects to the parameter estimation results. 

Considering the above motivations, in this section, we study the red noise
effects in the parameter estimation. In Section~\ref{subsec:red noise set up},
we describe our red noise model. The design matrix approximation is introduced
in Section~\ref{subsec:design matrix}. We present a modified Fisher matrix
analysis with the assumption that we know the red noise parameters in
Section~\ref{subsec:red bias}. We show results of Bayesian analysis that
estimate the system parameters and red noise parameters simultaneously in
Section~\ref{subsec:Bayesian approach}.

%----------------------------------------------------------------------
\subsection{Red noise model}\label{subsec:red noise set up}
%----------------------------------------------------------------------

The red noise component in the timing noise power spectrum density can be simply
described by a (two-sided) power-law spectrum,
\begin{equation}
	P_{\rm red}(f)=\frac{A_r^2}{12\pi^2}\left(\frac{f}{f_{\rm
	yr}}\right)^{-\alpha}\,,
\end{equation}
where $A_r$ is the red noise amplitude in the unit of ${\rm yr^{3/2}}$, $\alpha$
is the spectrum index, and $f_{\rm yr}=1\,{\rm yr^{-1}}$. However, for
simulating a finite noise realization and for the convenience in the later
analytical analysis without the need to deal with various divergence terms, we
adopt a cut-off power law spectrum~\citep{Coles:2011zs}
\begin{equation}\label{eq:cut-off PSD}
	P_{\rm red}(f)=\frac{A}{\big[1+(f/f_c)^2 \big]^{\alpha/2}}\,,
\end{equation}
where $f_c$ is a cut-off frequency for the low-frequency part, and 
\begin{equation}
	A=\frac{A_r^2}{12\pi^2}\left(\frac{f_c}{f_{\rm yr}}\right)^{-\alpha}\,.
\end{equation}
Due to the fact that one always fits the pulsar's rotation, characterized by
$N_0$, $\nu$ and $\dot{\nu}$, in the timing model, the spectrum will always be
flattened below the frequency $1/T_{\rm obs}$. For $f_c<1/T_{\rm obs}$, the
cut-off power-law spectrum is equivalent to the simple power law for our
purpose. Therefore, we fix $f_c=1/50\,{\rm yr^{-1}}$ in this paper. We have
verified that changing $f_c$ does not affect our results significantly.

%----------------------------------------------------------------------
\begin{figure*}[ht!]
	\begin{center}
		\includegraphics[width=17cm]{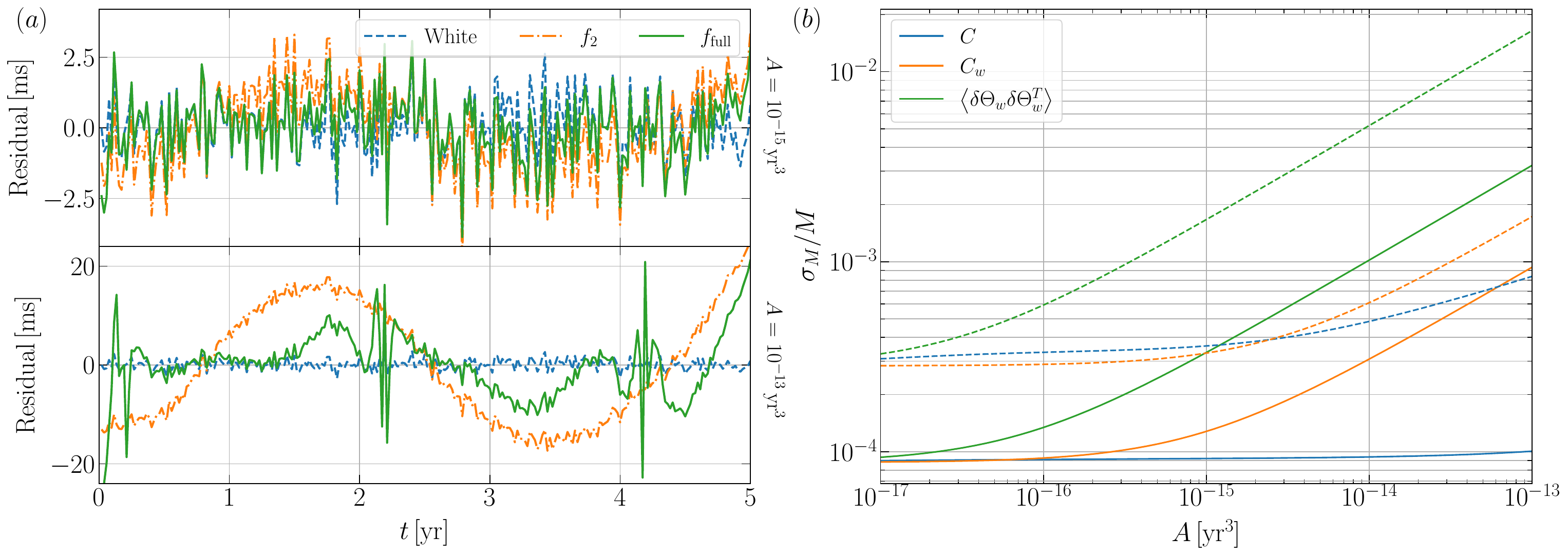}
	\caption{(a) Two noise realizations with different red noise amplitude $A$. The
	power-law index is chosen to be $\alpha=5$ for both panels. The white-noise
	components are shown by the blue dashed line and we have used $\sigma_{\rm
	TOA}=1\,{\rm ms}$. $f_2$ denotes the timing residuals after a second-order
	polynomial fitting that subtracts the pulsar rotation's contribution via
	$N_0$, $\nu$, and $\dot{\nu}$. $f_{\rm full}$ denotes the timing residuals
	after a fitting of the linearized full timing model. Note that the fitting
	here is a simple least-square fitting which does not consider noise
	properties. (b) The fractional precision of the mass of  SMBH as a function of the
	red noise amplitude estimated by three different methods shown in
	Equations~(\ref{eq:cov_correct}), (\ref{eq:cov_white_F}), and
	(\ref{eq:cov_white_T}). For the covariance matrix $C_w$ we use the noise
	averaged $\sigma^2_{w,{\rm eff}}$ shown in
	Equation~(\ref{eq:sigma_eff_ave}). The solid line shows the results for a
	pulsar-SMBH system with system parameters same as in Section~\ref{sec:timing
	model}, while the dashed line shows a similar system but with a larger
	orbital period $P_b=2\,{\rm yr}$. \label{fig:noise_realization}}
	\end{center}
\end{figure*}
%----------------------------------------------------------------------

Figure~\ref{fig:noise_realization}~(a) shows two noise realizations, including both
red noise and white noise, but with different red noise amplitudes. For the
white-noise component, we consider $\sigma_{\rm TOA}=1\,{\rm ms}$ as used in
previous sections, and the realizations are shown by the blue-dotted line. We
also show the post-fit residuals of the noise realizations. In the figure $f_2$
denotes the timing residuals after a second-order polynomial fitting that
subtracts the pulsar's rotation contribution (via parameters $N_0$, $\nu$, and
$\dot{\nu}$); $f_{\rm full}$ denotes the timing residual after a fitting of the
linearized full timing model that is introduced later in this subsection. For
the linearized full timing model, the system parameters are described in
Section~\ref{sec:timing model} except that here we use $P_b=2\,{\rm yr}$. The
clear absorption of the timing residuals by the linearized full timing model
suggests the existence of strong biases in system parameters besides $N_0$,
$\nu$, and $\dot{\nu}$.

The amplitude $A$ and power-law index $\alpha$ can vary a lot for different
pulsars. Here we choose the values of $A$ and $\alpha$ motivated by the studies
of a large sample of pulsars~\citep{Shannon:2010bv,Parthasarathy:2019txt}. For
normal pulsars, the spectrum index $\alpha$ peaks around $5$, and the red noise
amplitude  varies around $A_{r}\sim 10^{-10}\,{\rm yr^{-3/2}}$ for more than one
dex. Transforming this value to $A$ as a reference, in
Figure~\ref{fig:noise_realization}, we choose $A=10^{-15}\,{\rm yr^3}$ to
represent a case where the red noise is relatively weak and comparable to the
white noise, and a case where $A=10^{-13}\,{\rm yr^3}$ which represents a
situation where the pulsar has a strong red noise dominating. 

In this work, we ignore the red noise contribution caused by the dispersion
measure variation. Though, as mentioned before, the complex astrophysical
environment in the GC region may introduce a relatively large red noise
contribution, they can be treated similarly to the pulsar's intrinsic red noise
in real data analysis. Including it will only introduce another power-law
component in the noise spectrum~\citep{Chen:2025qqb}. Further, the contribution
from dispersion measure variation has an additional dependence on the
observation frequency as $P_{\rm DM}\propto \nu_{\rm obs}^{-4}$. \ZX{}{Future observation 
at high radio frequencies will further suppress this contribution, and it may be 
negligible unless the ISM around the GC is especially turbulent.}

%----------------------------------------------------------------------
\subsection{Design matrix approximation}
\label{subsec:design matrix}
%----------------------------------------------------------------------

Discussions in Section~\ref{sec:parameter estimation} used simulated
data sets\ZX{}{with zero-noise realizations} for prospecting the measurement precision of various system
parameters. In the presence of noise, in principle, one shall use the fully
non-linear timing model constructed in Section~\ref{sec:timing model} to fit the
TOAs and find the best-fit parameters. In this investigation, as we are only
aiming to study the possible effects caused by the presence of red noise, we
adopt the simplification of the design matrix. That is, one may expand the
non-linear timing model around a system parameter set that is close enough to
the best-fit point, and only takes the linear order of the timing model that is
described by the design matrix,
\begin{equation}
	\mathcal{M}_{i\mu}=-\frac{1}{\nu}\frac{\partial
	\mathcal{N}_i}{\partial\Theta^\mu}\,.
\end{equation}
For real observations, the expansion point can be obtained by, for example, a
primary fit. Here we simply expand the timing model around the  system's true
parameter set $\bar{\Theta}$, which is valid as long as the noise is small. This
choice is also equivalent to the Fisher matrix approximation in the case of pure
white noise, as the covariance matrix in the Fisher matrix approximation defined
by Equation~(\ref{eq:cov matrix2}) has the form 
\begin{equation}
	C_{\alpha\beta}=\left(\sum_{i}\frac{1}{\sigma^2_{ {\rm
	TOA},i}}\mathcal{M}_{i\alpha} \mathcal{M}_{i\beta}\right)^{-1}\,.
\end{equation}

With the above assumptions, the timing residuals in the presence of noise are 
\begin{eqnarray}\label{eq:resi_M}
	\mathcal{R}_i &=& \frac{1}{\nu}\Big[\mathcal{N} \big(t^{\rm TOA}_i;\Theta \big)
	-\mathcal{N} \big(t^{\rm TOA}_i;\bar{\Theta} \big)\Big]\,,\nonumber\\
	&=&-\mathcal{M}_{i\mu}\delta\Theta^\mu+\delta t_i\,,
\end{eqnarray}
where $\delta\Theta^\mu=\Theta^\mu-\bar{\Theta}^\mu$ as we expand the timing
model around the  system's true parameter set $\bar{\Theta}$. In
Equation~(\ref{eq:resi_M}) a summation on $\mu$ is performed according to the
Einstein summation convention. $\delta t_i$ represents the pre-fit timing
residuals\footnote{It is in fact the timing residuals after the primary fit,
which gives $\bar{\Theta}$ in our assumption.} and can be simulated by the noise
model we discussed before. For the \ZX{}{zero-noise realization} case, $\delta t_i=0$, and it is
clear that the best-fit parameters are the system's true parameters, as we do
not consider mismodeling effects here. However, in the presence of noise,
maximizing a probability like Equation~(\ref{eq:probability}) will lead to a
non-zero $\delta\Theta^\mu$ which depends on the noise realization.

%----------------------------------------------------------------------
\subsection{Bias on parameter estimation}\label{subsec:red bias}
%----------------------------------------------------------------------

In the presence of red noise, the probability function is no longer described by
Equation~(\ref{eq:probability}). Instead, we have 
\begin{equation}\label{eq:porb_C}
	P(\Theta|{t^{\rm TOA}})\propto\exp\left(-\frac{1}{2}
	\mathcal{R}^T\mathcal{C}^{-1} \mathcal{R}\right)\,.
\end{equation}
Here and in the following, we use the matrix notation so that $\mathcal{R}$ is
the column vector constructed by $N_{\rm TOA}$ timing residuals, and
$\mathcal{C}$ is the $N_{\rm TOA} \times N_{\rm TOA}$ noise covariance matrix
defined by 
\begin{equation}
	\mathcal{C}=\mathcal{C}_{w}+\mathcal{C}_{r}\,.
\end{equation}
For the white noise contribution, we simply consider the form 
\begin{equation}\label{eq:noise_cov_w}
	\mathcal{C}_{w,ij}=\sigma^2_{\rm TOA}\delta_{ij}\,,
\end{equation}
with $\delta_{ij}$ the Kronecker delta symbol, though for real observations and
analysis it can be more complex~\citep{Chen:2025qqb}. The red noise contribution
is, by definition, the inverse Fourier transformation of the power spectral
density, which reads
\begin{equation}
	\mathcal{C}_{r,ij}=\left<\delta t_i\delta t_j\right> =\int_{-\infty}^\infty
	P_{\rm red}(|f|) e^{i2\pi f\tau_{ij}}{\rm d}f\,,
\end{equation}
where $\left<\cdot\right>$ is the average over noise realizations, and
$\tau_{ij}=t^{\rm TOA}_i-t^{\rm TOA}_j$. Note that here we use the convention
that $P_{\rm red}(f)$ is the two-sided power spectrum. For $P_{\rm red}(f)$ in
Equation~(\ref{eq:cut-off PSD}), this is 
\begin{equation}
	\mathcal{C}_{r,ij}=\frac{2\pi^{\alpha/2}A}{\Gamma(\alpha/2)}
	f_c^{\frac{\alpha+1}{2}}
	|\tau_{ij}|^{\frac{\alpha-1}{2}}K_{\frac{\alpha-1}{2}} \big(2\pi
	f_c|\tau_{ij}| \big)\,,
\end{equation}
where $K_n(z)$ is the modified Bessel function of the second kind and
$\Gamma(z)$ is the Gamma function. For $i=j$ one has
\begin{equation}
	\mathcal{C}_{r,ii}=\pi^{1/2}Af_c
	\frac{\Gamma((\alpha-1)/2)}{\Gamma(\alpha/2)}\,.
\end{equation}

If one knows exactly the power spectrum of the red noise and white noise,
therefore the noise covariance matrix $\mathcal{C}$, one can simply maximize the
probability function in Equation~(\ref{eq:porb_C}) within the linearized timing
model and Equation~(\ref{eq:resi_M}) via~\citep{Coles:2011zs}
\begin{equation}\label{eq:best_fit_red}
	\delta\Theta=\left(\mathcal{M}^{T}\mathcal{C}^{-1}
	\mathcal{M}\right)^{-1}\mathcal{M}^{T} \mathcal{C}^{-1}\delta t\,,
\end{equation}
for a given noise realization $\delta t$. The parameter estimation uncertainties
and correlations are given by the covariance matrix
\begin{equation}\label{eq:cov_correct}
	C=\left(\mathcal{M}^{T}\mathcal{C}^{-1}\mathcal{M}\right)^{-1}\,.
\end{equation}

However, in real observations, the noise parameters are {\it a priori} unknown.
Estimation of the noise parameters with methods like the generalized
least-square approach~\citep{Coles:2011zs} or Bayesian
approach~\citep{Chen:2025qqb} is needed and will be discussed in the next
subsection. In the rest of this subsection, we aim to make an intuitive
estimation of the biases caused by ignoring the red noise effects under the
assumed red noise model, which highlights the importance of performing detailed
noise analysis in pulsar-SMBH systems. 

Using the probability function defined in Equation~(\ref{eq:probability})
assumes white noise that has the same strength for all TOAs. If the assumption
holds, the post-fit residuals will have a $\chi^2$ distribution with $N_{\rm
TOA}-N_{\rm param}$ degrees of freedom, where $N_{\rm param}$ is the number of
parameters that were used in the fitting. Therefore, from observational data,
one may estimate the strength of the white noise via
\begin{equation}
	\sigma^2_{w,{\rm eff}}=\frac{\mathcal{R}^{T}_w \mathcal{R}_w}{N_{\rm
	TOA}-N_{\rm param}}\,,
\end{equation}
where $\sigma_{w,{\rm eff}}$ is the effective timing precision estimated from
data assuming purely white noise, and 
\begin{equation}
	\mathcal{R}_w=\delta t -\mathcal{M}\left(\mathcal{M}^{T}
	\mathcal{M}\right)^{-1}\mathcal{M}^{T} \delta t\,,
\end{equation}
is the post-fit timing residual. 

In the presence of red noise, one may still use the above estimation, which will
account for the red noise by an effective white noise $\sigma_{w,{\rm eff}}$.
The best-fit parameter in this estimation is unbiased in the sense that 
\begin{equation}
	\left<\delta\Theta_w\right> =\left(\mathcal{M}^{T}
	\mathcal{M}\right)^{-1}\mathcal{M}^{T} \left<\delta t\right>=0\,,
\end{equation}
where we use $\delta\Theta_w$ to distinguish it from the correct estimation
$\delta\Theta$ in Equation~(\ref{eq:best_fit_red}). However, the parameter
uncertainties estimated from the white noise assumption 
\begin{equation}\label{eq:cov_white_F}
	C_w=\left(\frac{1}{\sigma^2_{w,{\rm eff}}}
	\mathcal{M}^{T}\mathcal{M}\right)^{-1}\,,
\end{equation}
is different from the correct one given in Equation~(\ref{eq:cov_correct}).
Though by definition, $\sigma^2_{w,{\rm eff}}$ is estimated from the data and
can be different for each noise realization, for our purpose, it is sufficient
to use the noise-averaged value 
\begin{equation}\label{eq:sigma_eff_ave}
	\left<\sigma^2_{w,{\rm eff}}\right> =\frac{\left<\delta t^{T}
	\mathcal{Q}^T\mathcal{Q}\delta t\right>} {N_{\rm TOA}-N_{\rm param}}
	=\frac{{\rm Tr}\,(\mathcal{C} \mathcal{Q}^T \mathcal{Q} )}{N_{\rm
	TOA}-N_{\rm param}}\,,
\end{equation}
where 
\begin{equation}\label{eq:white_esti}
	\mathcal{Q}=\mathcal{I}-\mathcal{M}
	\left(\mathcal{M}^{T}\mathcal{M}\right)^{-1} \mathcal{M}^{T}\,,
\end{equation}
with $\mathcal{I}$ the identity matrix. Further, the estimation in
Equation~(\ref{eq:white_esti}) even does not represent the true covariance of
$\delta\Theta_w$ as
\begin{equation}\label{eq:cov_white_T}
	\left<\delta\Theta_w\delta\Theta_w^{T}\right> =\left(\mathcal{M}^{T}
	\mathcal{M}\right)^{-1} \mathcal{M}^{T}\mathcal{C} \mathcal{M}\left(
	\mathcal{M}^{T}\mathcal{M}\right)^{-1}\,,
\end{equation}
while one can easily verify that Equation~(\ref{eq:cov_correct}) is consistent
with Equation~(\ref{eq:best_fit_red}), and $C=\left<\delta\Theta\delta\Theta^{T}\right>$.  

In  Figure~\ref{fig:noise_realization}~(b), we show the estimated measurement precision of
the mass of  SMBH as a function of the red noise amplitude. We show the results
both from a correct estimation given by Equation~(\ref{eq:cov_correct}), which
requires the knowledge of the noise covariance matrix $\mathcal{C}$, and the
estimation based on the assumption of purely white noise as given in
Equation~(\ref{eq:cov_white_F}). However, as we mentioned above,
Equation~(\ref{eq:cov_white_F}) does not correctly give the uncertainty of the
best-fit parameter $\delta\Theta_w$. Therefore, we also show the true
uncertainty of $\delta\Theta_w$ given by Equation~(\ref{eq:cov_white_T}). The
solid line and dashed line in the figure show the results for two pulsar-SMBH
systems that have $P_b=0.5\,{\rm yr}$ and $P_b=2\,{\rm yr}$ respectively. Other
system parameters are the same as in Section~\ref{sec:timing model}. 

From the figure, one can see that, with full knowledge of the noise properties,
the parameter estimation uncertainty does not increase a lot even when the red
noise has a large amplitude, while the two estimations based on the white noise
assumption are largely affected by the presence of red noise. This is because
that by assuming a purely white noise, the noise power is incorrectly absorbed
by the timing model parameters, and leads to a larger deviation that depends on
the noise realization. 

More importantly, the figure suggests that assuming a purely white noise also
leads to a wrong estimation of the real parameter uncertainty. One can only use
Equation~(\ref{eq:cov_white_F}) as the estimator, as the correct estimation in
Equation~(\ref{eq:cov_white_T}) again requires the knowledge of $\mathcal{C}$.
From the figure, one can see that the estimation in
Equation~(\ref{eq:cov_white_F}) significantly underestimates the parameter
uncertainties, which might lead to a false alarm, for example, in deviation of
GR when testing gravity. Therefore, careful treatment of the parameter
estimation in the presence of red noise is necessary in  pulsar-SMBH systems. 

%----------------------------------------------------------------------
\subsection{Bayesian approach to parameter estimation}
\label{subsec:Bayesian approach}
%----------------------------------------------------------------------

The discussion above suggests that it is necessary to consider the possible red
noise in  pulsar-SMBH systems. However, in real observations, the noise
properties are {\it a priori} unknown and need to be estimated from the data. In
this subsection, we study the effects of red noise in parameter estimations for
pulsar-SMBH systems based on the Bayesian approach which is well developed and
widely used in pulsar timing~\citep{Lentati:2012xb, NANOGrav:2023gor,
Chen:2025qqb}. Such an approach can estimate the red noise parameters from the
timing data simultaneously with other parameters.

For convenience, we first introduce the approach mainly
following~\citet{Lentati:2012xb}. Based on the Bayes' theorem, the posterior for
the system parameters $\Theta$, the noise parameters $\Xi$, and a set of
auxiliary noise variables $\epsilon$ that will be introduced later, can be
written as 
\begin{eqnarray}
	P \big(\Theta,\Xi,\epsilon|t^{\rm TOA} \big) & \propto& P \big(\delta
	t|\delta \Theta,\Xi,\epsilon \big)P \big(\delta\Theta, \Xi,\epsilon \big)\\
	&\propto& P \big(\delta t|\delta\Theta,\epsilon\big)P \big(\epsilon|\Xi
	\big)P (\delta \Theta )P(\Xi)\,, \nonumber
\end{eqnarray}
where we have assumed that a primary fit has deduced $t^{\rm TOA}$ and $\Theta$
to $\delta t$ and $\delta \Theta$ respectively. The proportion symbol
``$\propto$'' used in this subsection exactly means a normalization factor. With
linear approximation, the probability function of the pre-fit residual $\delta
t$, given $\delta \Theta$ and $\epsilon$, reads
\begin{eqnarray}\label{eq:P_t_theta_epsilon}
	P(\delta t|\delta\Theta,\epsilon) &\propto&\frac{\exp
	\left[-\frac{1}{2}(\mathcal{R} -\mathcal{F} \epsilon)^{T}C_w^{-1}
	(\mathcal{R}-\mathcal{F}\epsilon) \right]}{\sqrt{ \det{\mathcal{C}_w}}}\,,
\end{eqnarray}
with $\mathcal{R}=\delta t-\mathcal{M}\delta\Theta$ given in
Equation~(\ref{eq:resi_M}). The matrix $\mathcal{F}$ is a set of Fourier bases
that aim to account for the red noise component in the pre-fit residual $\delta
t$, and therefore $\epsilon$ gives the mode amplitude. We set 
\begin{eqnarray}
	\mathcal{F}_{i,l}& =&\cos \big(2\pi t^{\rm TOA}_i l/T_{\rm obs}
	\big)\,,\nonumber\\
	\mathcal{F}_{i,k+l}&=&\sin \big(2\pi t^{\rm TOA}_i l/T_{\rm obs}\big)\,,
\end{eqnarray}
with $i=1,\cdots,N_{\rm TOA}$, and $l=1,\cdots,k$. The number $k$, as a
hyperparameter, determines the frequency band that we want to remove the red
noise, and its optimal value will be determined later. Finally, $C_w$ in
Equation~(\ref{eq:P_t_theta_epsilon}) is the noise covariance matrix of purely
white noise, as we have explicitly subtracted the red noise component by
$\mathcal{F}\epsilon$. As per convention~\citep{Chen:2025qqb}, we add a
multiplicative factor ${\rm EFAC}$ to Equation~(\ref{eq:noise_cov_w}) as 
\begin{equation}
	\mathcal{C}_w={\rm EFAC}^2\sigma_{\rm TOA}^2\mathcal{I}\,,
\end{equation}
and fix $\sigma_{\rm TOA}^2=1\,{\rm ms}$. For real date analysis, one may
similarly include all three factors, i.e., $\rm EFAC$, $\rm EQUAD$, and $\rm
ECORR$.

$P \big(\epsilon|\Xi \big)$ describes the probability of the amplitudes
$\epsilon$ given the noise parameters. Therefore we have
\begin{equation}
	P \big(\epsilon|\Xi \big)\propto \frac{1}{\sqrt{\det{\varphi}}}
	\exp\left(-\frac{1}{2}\epsilon^T \varphi^{-1}\epsilon\right)\,,
\end{equation}
where
\begin{equation}\label{eq:varphi}
	\varphi_{ij}=\left<\epsilon_i\epsilon_j\right>=\frac{2}{T_{\rm
	obs}}P_r(f_i|\Xi)\delta_{ij}\,.
\end{equation}
The function $P_r$ is the red noise spectrum that depends on the noise
parameters $\Xi$, and 
\begin{equation}
	f_i=f_{i+k}=i/T_{\rm obs}\,,
\end{equation}
is the corresponding frequency. Note that the factor ``$2$'' in
Equation~(\ref{eq:varphi}) comes from the convention that we are using the
two-sided power spectrum. 

According to the power-law model of the red noise, we choose
\begin{equation}
	P_r(f|\Xi)=\tilde{A}T_{\rm obs}^{3} \left(fT_{\rm obs}\right)^{-\alpha}\,,
\end{equation} 
where $\tilde{A}$ is the dimensionless amplitude defined by $\tilde{A}=AT_{\rm
obs}^{-3}\left(f_cT_{\rm obs}\right)^{\alpha}$. We treat both $\tilde{A}$ and
$\alpha$ as noise parameters. Together with $\rm EFAC$ we have
\begin{equation}
	\Xi= \big\{{\rm EFAC},\tilde{A},\alpha \big\}\,.
\end{equation}

We shall point out that, if one adopts the red-noise model with a low frequency
cutoff as in Equation~(\ref{eq:cut-off PSD}), one may not introduce the Fourier
components and can directly model the red noise effects with
$\mathcal{C}_r$~\citep{vanHaasteren:2012hj}. The two methods should provide
consistent results. The approach we adopted here has the advantage that we do
not need to model the red noise spectrum for low frequencies.

The priors of $\delta\Theta$ and $\Xi$ are given by $P(\delta\Theta)$ and
$P(\Xi)$ respectively. As we use the linear approximation, it is natural to
choose $P(\delta\Theta)$ to be the flat prior, and we will not discuss it from
now on. For the noise parameters, we set ${\rm EFAC}\sim\mathcal{U}(0.1,10)$,
$\log_{10}\tilde{A}\sim \mathcal{U}(-30,-10)$, and $\alpha\sim\mathcal{U}(0,10)$
with $\mathcal{U}$ denoting the uniform distribution~\citep{Chen:2025qqb}. 

As studied by \citet{Lentati:2012xb} and \citet{vanHaasteren:2012hj}, for
estimating the red noise parameter $\Xi$, one can analytically marginalize over
the Fourier amplitudes $\epsilon$ and the timing parameters $\delta\Theta$ when
under the assumption of a flat prior. Integrating over $\epsilon$ gives
\begin{equation}\label{eq:posterior_theta_xi}
	P(\delta\Theta,\Xi|\delta t) \propto
	P(\Xi)\frac{\exp\Big[-\frac{1}{2}\big(\mathcal{R}^{T}
	\mathcal{C}_w^{-1}\mathcal{R}-d^T\Sigma^{-1}d\big)
	\Big]}{\sqrt{\det{\varphi} \det{\mathcal{C}_w}\det{\Sigma}}}\,,
\end{equation}
where
\begin{eqnarray}
	d&=&\mathcal{F}^{T}\mathcal{C}^{-1}_w\mathcal{R}\,,\\
	\Sigma&=&\mathcal{F}^{T}\mathcal{C}_w^{-1} \mathcal{F}+\varphi^{-1}\,.
\end{eqnarray}
Further marginalizing $\delta\Theta$ gives us a simple formula that can be used
to directly sample the posterior of the noise parameters,
\begin{equation}\label{eq:likelihood_Xi}
	P(\Xi|\delta t)\propto P(\Xi)\frac{\exp\left(-\frac{1}{2}\delta t^T
	\mathcal{X} \delta t\right)}{\sqrt{\det{\varphi}
	\det{\mathcal{C}_w}\det{\Sigma}\det(\mathcal{M}^{T}\mathcal{Y}\mathcal{M})}}\,,
\end{equation}
where
\begin{eqnarray}
	\mathcal{Y}&=&\mathcal{C}_w^{-1}- \mathcal{C}_w^{-1}\mathcal{F}
	\Sigma^{-1}\mathcal{F}^{T} \mathcal{C}_w^{-1}\,,\\
	\mathcal{X}&=&\mathcal{Y}-\mathcal{Y} \mathcal{M}(\mathcal{M}^{T}
	\mathcal{Y}\mathcal{M})^{-1} \mathcal{M}^{T}\mathcal{Y}\,.
\end{eqnarray}
For numerical stability, one may first perform a singular value decomposition on
the design matrix such that $\mathcal{M}=U\mathcal{S}V^{T}$.\footnote{Here we
adopt the convention that $\mathcal{S}$ and $V$ are square matrices.} Then one
has
\begin{equation}
	\mathcal{X}=\mathcal{Y}-\mathcal{Y}U(U^{T}
	\mathcal{Y}U)^{-1}U^{T}\mathcal{Y}\,,
\end{equation}
and
\begin{equation}
	\det(\mathcal{M}^{T}\mathcal{Y}\mathcal{M})= \det(U^{T}\mathcal{Y}U)
	\det\mathcal{S}^2\,.
\end{equation}
Note that the term $\det\mathcal{S}^2$ can be dropped as it is a constant that
does not depend on $\Xi$.

%----------------------------------------------------------------------
\begin{figure}[t]
	\begin{center}
		\includegraphics[width=8.6cm]{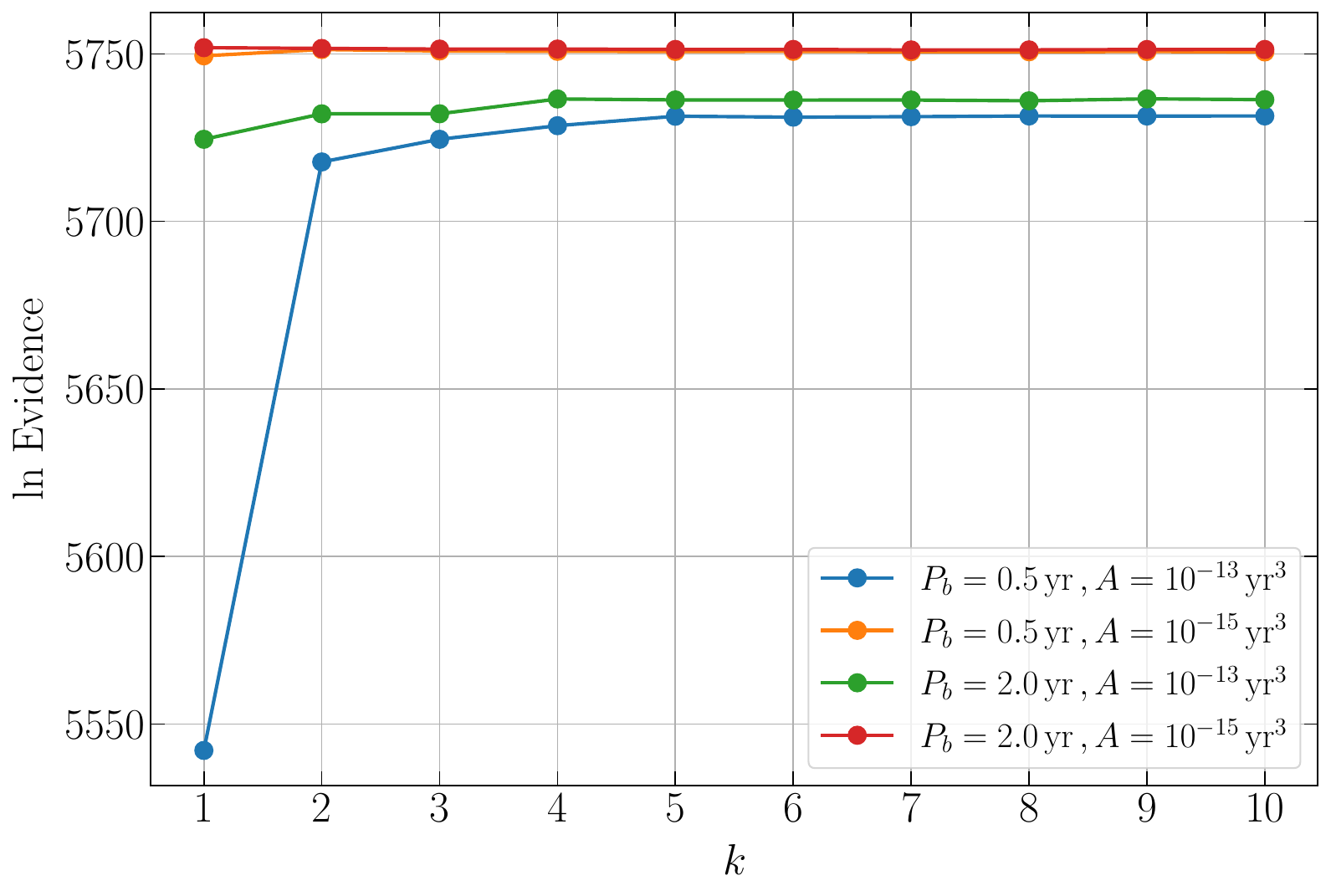}
	\caption{The evidence for four examples as functions of the Fourier bases
	number $k$.  \label{fig:evidence}}
	\end{center}
\end{figure}
%----------------------------------------------------------------------

With the above formulae, and the same noise injection shown in
Figure~\ref{fig:noise_realization}, we perform nested sampling parameter
estimation with the \texttt{dynesty} sampler in the \texttt{BILBY}
package~\citep{Ashton:2018jfp} to obtain the samples of the posterior
$P(\Xi|\delta t)$. We consider four cases that have pulsar's orbital periods
$P_b=0.5\,{\rm yr}$ or $P_b=2.0\,{\rm yr}$, and red noise amplitude
$A=10^{-13}\,{\rm yr^3}$ or $A=10^{-15}\,{\rm yr^3}$. For all these cases, we
perform $10$ samplings with $k$ varying from 1 to 10. The evidences estimated by
the sampler are shown in Figure~\ref{fig:evidence}. From the figure, one can see
that, for the stronger red noise, $A=10^{-13}\,{\rm yr^3}$, the evidences
increase before about $k=5$ for both cases, which suggests that including more
Fourier components better models the noise. For the cases where the red noise is
comparable to the white noise, the increase in $k$ does not significantly change
the evidence, which implies that it will be hard to estimate the red noise
parameters, especially the spectrum index $\alpha$. One can also notice that,
when the pulsar has a larger orbital period, the change in evidence caused by
the modeling of red noise is smaller, as we expect that the timing model will
absorb more red noise power when the pulsar's orbital period is larger.  

%----------------------------------------------------------------------
\begin{figure*}[ht!]
	\begin{center}
	\includegraphics[width=16cm]{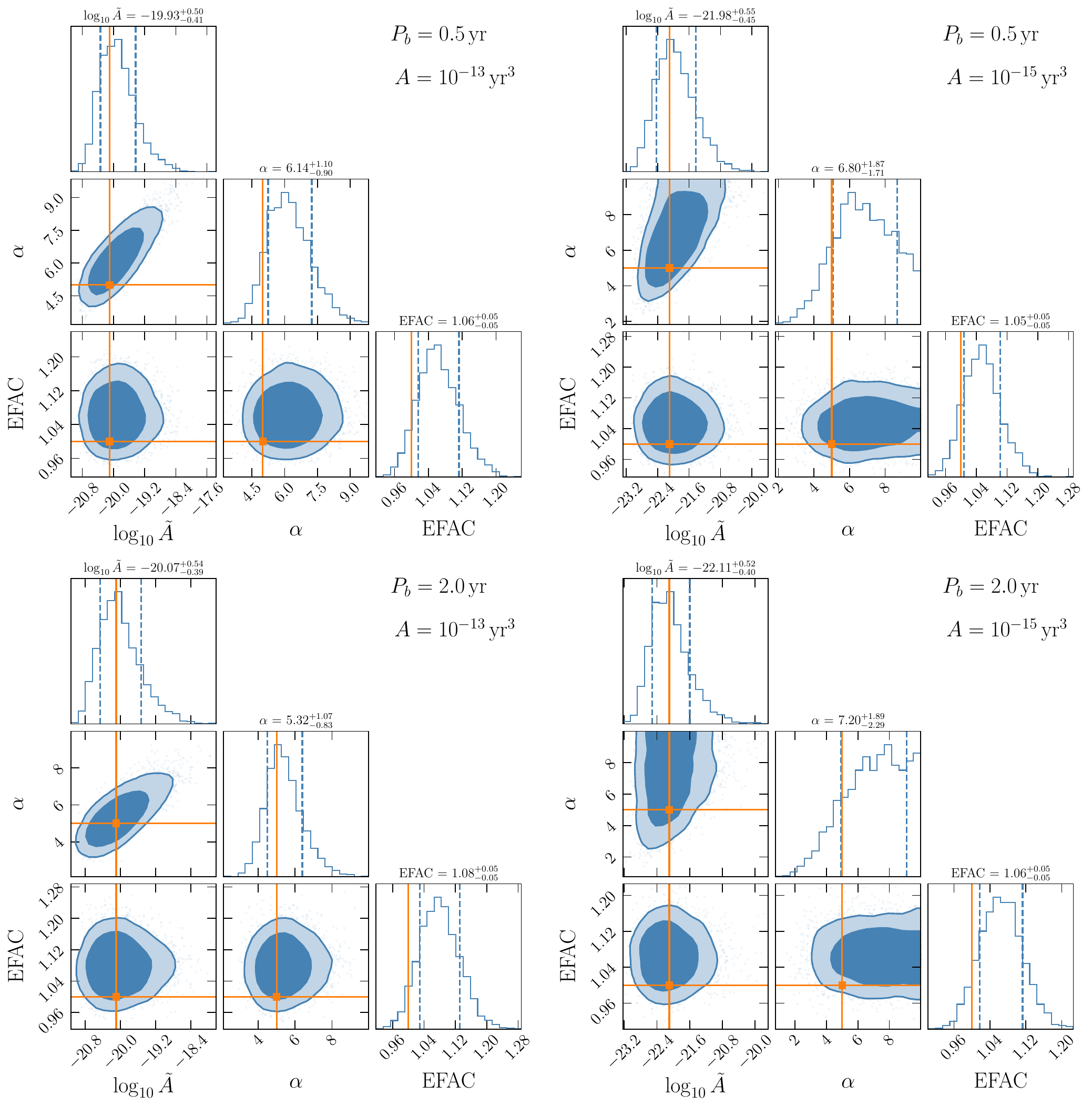}
	\caption{Posteriors of red-noise parameter for four cases with different
	pulsar orbital periods and red noise amplitudes. The number of Fourier bases
	is $k=6$ for all these figures. The true values of the parameters are marked
	by the orange lines and squares. The dashed lines show the 1-$\sigma$
	intervals of the 1-D marginalized distributions, while the contours show the
	$68\%$ and $90\%$ credible regions in the 2-D parameter space.
	\label{fig:noise_para_4case}}
	\end{center}
\end{figure*}
%----------------------------------------------------------------------

Figure~\ref{fig:noise_para_4case} shows the posteriors of the red-noise
parameters for the four examples. For all cases and later analysis, we choose
$k=6$ according to previous discussions. But we note that $k$ may vary in other
cases and needs to be estimated as above. From the figure, one can see that, as
expected, when the red noise is strong, one can extract the red noise parameters
from the timing residuals. But for weak noise cases, though the red noise
amplitude can be estimated, the spectrum indices are harder to measure,
especially for short observation time spans, similar to the case of measuring
the stochastic gravitational wave background~\citep{Janssen:2014dka}. 

%----------------------------------------------------------------------
\begin{figure*}[t]
	\begin{center}
		\includegraphics[width=17cm]{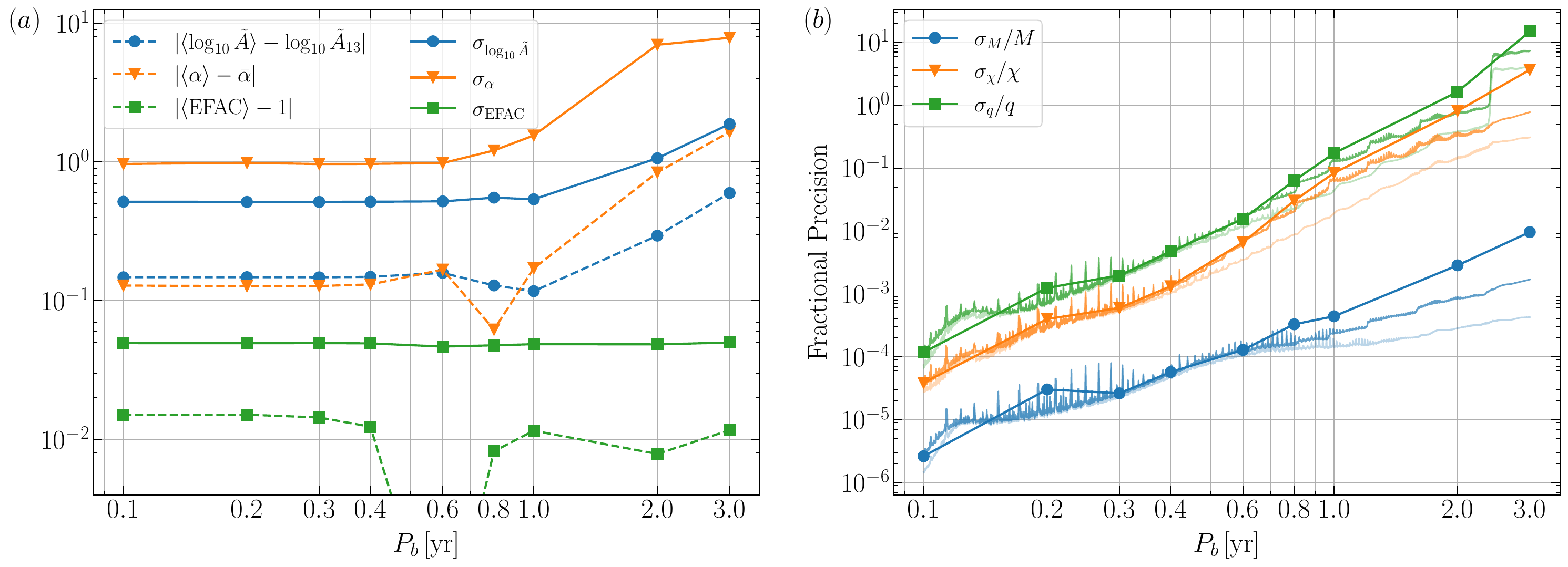}
	\caption{(a) The absolute difference between the averaged values of the
	best-fit noise parameters and true parameters, as well as the standard
	deviation of the best-fit parameters for $N_r=10000$ noise realizations. In
	this figure we have used $\tilde{A}_{13}=8\times10^{-21}$ and
	$\bar{\alpha}=5$. (b) The fractional precision of SMBH parameters as
	functions of the pulsar orbital period in the presence of red noise. We have
	used $A=10^{-13}\,{\rm yr^3}$ and $\alpha=5$. The points are the variance of
	the best-fit parameters in $N_r=10000$ noise realizations, while the fainter
	lines are the parameter estimation results  assuming that we know the true
	noise parameters or do not consider the red noise (see the same line in
	Figure~\ref{fig:full_PE-pm_Pb} (b)).  \label{fig:noise_statistic_13}}
	\end{center}
\end{figure*}
%----------------------------------------------------------------------

To better study the statistical property of the red noise estimation, we
generate $N_r=10000$ noise realizations for $A=10^{-13}\,{\rm yr^3}$. Limited by
the computational resources, instead of performing the Bayesian analysis, we
directly minimize the likelihood function $P(\Xi|\delta t)/P(\Xi)$ based on
Equation~(\ref{eq:likelihood_Xi}) to obtain the best-fit values $\Xi_b$ of the
red noise parameters. Note that minimizing this likelihood is assuming a
generalized flat prior instead of the uniform distribution in a finite range
that we used before. Nevertheless, they are the same as long as the best-fit
point is inside the prior range. 

In Figure~\ref{fig:noise_statistic_13}~(a), we show the mean and variance of the
best-fit parameters as functions of the pulsar orbital period for the
$A=10^{-13}\,{\rm yr^3}$ case. As shown in the figure, for each parameter, the
deviation of the mean from its true value is well below the variance, which
suggests that the parameter estimation result is practically unbiased. Further,
one can see that the variances of the two red noise parameters are increasing
for $P_b\gtrsim 1\,{\rm yr}$, which is consistent with our previous discussion.
Also, the variance calculated here can be regarded as an estimation of the
measurement precision of the red noise parameters.

%----------------------------------------------------------------------
\begin{figure}[t]
	\begin{center}
	\includegraphics[width=8.6cm]{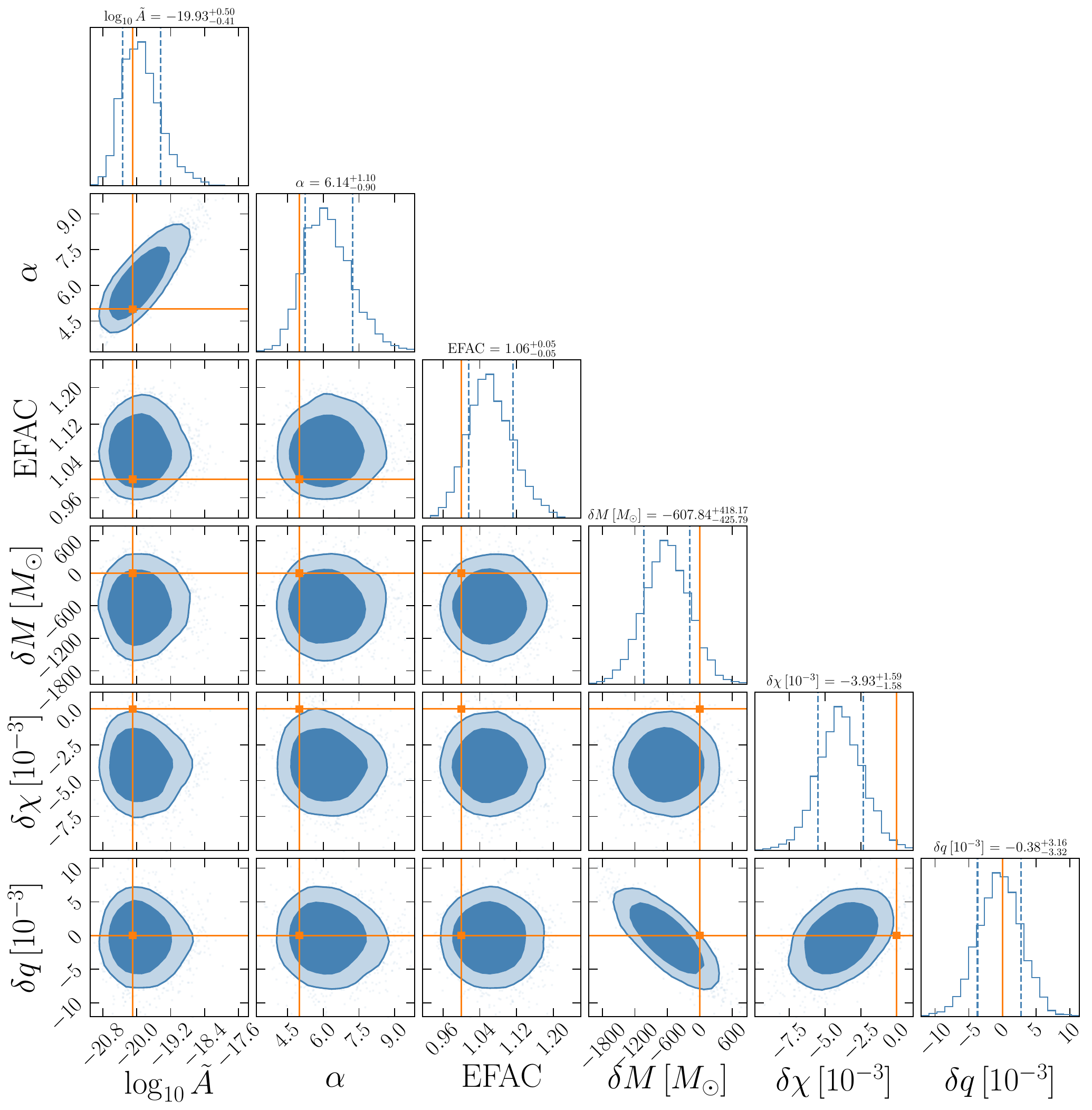}
	\caption{The recovered posterior for the $P_b=0.5\,{\rm yr}$ and
	$A=10^{-13}\,{\rm yr^3}$ case shown before. We only show the posteriors of
	the noise parameters, and $M$, $\chi$, $q$ of the SMBH. No clear correlation
	between noise parameters and timing parameters is found. \label{fig:recover_Pb_0.5_Ae13}}
	\end{center}
\end{figure}
%----------------------------------------------------------------------

After estimating the red noise parameters, one can easily recover the posteriors
of the full parameter set using the resampling technique, which is widely used
in many areas~\citep[see e.g.,][]{Thrane:2018qnx,Dong:2025igh}. To recover the
posterior of $\delta\Theta$ and $\Xi$ in Equation~(\ref{eq:posterior_theta_xi}),
one may simply draw a corresponding sample of $\delta \Theta$ for each $\Xi$
sample obtained in the previous nested sampling process. The probability
function for the resampling is
\begin{eqnarray}
	&&P \big(\delta\Theta|\Xi,\delta t\big )\propto\nonumber\\ 
	&&\exp\left[-\frac{1}{2}\left(\delta\Theta-
	\delta\Theta_b\right)^T\mathcal{M}^{T} \mathcal{Y}
	\mathcal{M}\left(\delta\Theta- \delta\Theta_b\right)\right]\,,
\end{eqnarray}
where
\begin{equation}\label{eq:best_fit_timing_para}
	\delta\Theta_b=\left(\mathcal{M}^T\mathcal{Y}
	\mathcal{M}\right)^{-1}\mathcal{M}^T \mathcal{Y}\delta t\,,
\end{equation}
gives the best-fit values of the timing parameters.
Figure~\ref{fig:recover_Pb_0.5_Ae13} shows the recovered posterior of the noise
parameters as well as the $M$, $\chi$, $q$ parameters of the SMBH. We find no
clear correlation between the noise parameters and the timing parameters,
including those that are not shown in the figure. Nevertheless, the above
discussions show a practical approach that can estimate the noise parameters and
timing parameters simultaneously.

As before, to better understand the statistical properties of the red-noise
effects in the parameter estimation of  pulsar-SMBH systems, we obtain the
best-fit timing parameters for  $N_r=10000$ noise realizations based on
Equation~(\ref{eq:best_fit_timing_para}) and the corresponding best-fit noise
parameters. In  Figure~\ref{fig:noise_statistic_13}~(b), we show the fractional precision
of the SMBH parameters as functions of the pulsar orbital period in the presence
of red noise. For comparison, we show the results for three cases: (i) estimate
the timing parameters and noise parameters simultaneously; (ii) estimate the
timing noise only and assume that we know the true  parameters of red noise;
(iii) absence of red noise as discussed in Section~\ref{sec:parameter
estimation}. One can clearly see that the presence of red noise will increase
the parameter estimation uncertainty, especially for pulsars in larger orbits.
As expected, the curve represents the case in which we know the red noise
parameters are between two case (i) and case (iii). However, even for a red
noise that is relatively large, for example $A=10^{-13}\,{\rm yr^3}$ considered
in this figure, if one treats it properly, the increase of the measurement
precision is mild, especially for pulsars with orbital period $P_b\lesssim
T_{\rm obs}/10$. In contrast, improper treatments, as shown in
Figure~\ref{fig:noise_realization}~(b), may cause larger biases in the results of
parameter estimation. 

We shall also note that, in the presence of a red noise with a shallower
spectrum, for example, red noise caused by the DM variation, the estimation of
timing parameters for pulsars in smaller orbits might also be affected, as more
noise power will leak into the frequency of orbital motion. 

Finally, another common way of dealing with timing red noise is to include
higher-order spin derivatives in the pulsar's rotation model~(\ref{eq:proper
rotation}). For example, in the data analysis of the double pulsar, up to fourth
order spin derivatives were considered~\citep{Kramer:2021jcw}. In principle, the
higher-order spin derivatives will absorb part of the red noise, as in the
frequency domain, fitting these parameters effectively applies a high-pass
filter to the power spectrum of the timing residual. Nevertheless, we still
suggest to perform full noise analysis for pulsar-SMBH systems when possible.

%----------------------------------------------------------------------
\section{Conclusions} \label{sec:conclusions}
%----------------------------------------------------------------------

In this work, we construct a realistic pulsar-SMBH timing model based on
numerically integrating the PN equations of motion. We give a comprehensive
discussion of all the effects that need to be taken into account in the pulsar's
orbital motion and light propagation, including several next-to-leading-order
effects and the proper motion of Sgr~A*, which were not considered in previous
studies \citep[except for][which adopted a fully GR approach that intrinsically
includes all the effects but less extensible]{Zhang:2017qbb}.

Based on our timing model, we are able to first discuss the effects of the
proper motion of Sgr~A*. We find that the current VLBA observation already
provides a proper-motion measurement that is \ZX{}{more} accurate than what is possible from
timing measurement, which suggests that we can treat the proper motion of Sgr~A*
as fixed parameters when performing parameter estimation. This enables us to
measure the longitude of the ascending node $\Omega$ of the pulsar's orbit.
However, we find that the constraint on $\Omega$ is rather weak, especially for
pulsars with small orbits. Therefore, taking into account the proper motion of
Sgr~A* might not significantly break the leading-order degeneracy in spin
measurement, which is in contrast to the prospects in \citet{Zhang:2017qbb}. 

We also for the first time discuss the aberration effects in such systems. The
aberration effects are related to the spin of the pulsar. Due to the relatively
long spin-precession time scale in these systems, we treat the pulsar's spin
vector as constant in our investigation. We give the expression of the
next-to-leading-order aberration time delay and find that it might be relevant
for pulsars in compact orbits. According to our parameter estimation result,
based on timing-only observation, one might be able to constrain the direction
of the pulsar's rotation axis to $\sim1\,{\rm rad}$ level for pulsars with
orbital periods $P_b\lesssim 0.5\,{\rm yr}$.

Using this comprehensive timing model, we give a detailed discussion of the
parameter estimation for pulsar-SMBH systems. We consider the parameter
estimation in the presence of red noise. Differently from usual binary pulsar
systems, the red noise may affect the estimation of the orbital parameters due
to the large orbital timing scale of  pulsar-SMBH systems. Our results suggest
that, even for relatively strong red noise, treating it properly only leads to a
mild increase in the uncertainties of parameters, especially for pulsars with
orbital periods $P_b\lesssim T_{\rm obs}/10$. However, improper treatment, like
counting the red noise as effective white noise, may lead to significant biases
for parameter estimation. Worth to note that, in the current study, we
only consider a red noise with spectrum index $\alpha=5$, which is typical for
the intrinsic red noise of normal pulsars. However, red noise with a shallower
spectrum, for example, red noise caused by unmodeled dispersion-measure
variation, may cause a larger influence for pulsars with shorter orbital
periods. 

%----------------------------------------------------------------------
\begin{acknowledgments}
We thank Norbert Wex for helpful discussions, and Matthew Bailes for raising the
question about red noise. \ZX{}{We thank the anonymous referee for valuable comments.} This work was supported by the National Natural
Science Foundation of China (124B2056, 123B2043, 12573042), the National SKA
Program of China (2020SKA0120300), the Beijing Natural Science Foundation
(1242018), the Max Planck Partner Group Program funded by the Max Planck
Society, and the High-performance Computing Platform of Peking University.
\end{acknowledgments}
%----------------------------------------------------------------------
\appendix

%----------------------------------------------------------------------
\section{Next-to-leading-order aberration delay}
\label{app:second order aberration}
%----------------------------------------------------------------------

Here, we give a brief derivation of the aberration delay that is fully caused by
the orbital motion of the pulsar and special relativity to the next-to-leading
order. Consider that at the instant when the pulsar emits a pulse towards the
Earth, the pulsar has an orbital velocity $\bm{v}$. The light emitted by the
pulsar in the direction of $-\hat{\bm{K}}'_0$ seen in the proper frame of the
pulsar needs to point towards $-\hat{\bm{K}}_0$ in the center of mass frame of
the pulsar-SMBH system in order to reach the Earth. The velocity transformation
formula in special relativity thus gives us
\begin{eqnarray}
	-\hat{\bm{K}}'_0&=&-\hat{\bm{K}}_0-\left(\frac{\hat{\bm{K}}_0\cdot\hat{\bm{v}}
	+v/c}{1+\hat{\bm{K}}_0\cdot\bm{v}/c}-\hat{\bm{K}}_0
	\cdot\hat{\bm{v}}\right)\hat{\bm{v}}-[\hat{\bm{K}}_0-(\hat{\bm{K}}_0\cdot\hat{
	\bm{v}})\hat{\bm{v}}]\left(\frac{\sqrt{1-v^2/c^2}}{1+\hat{\bm{K}}_0\cdot\bm{v}
	/c}-1\right)\nonumber\\
	&=&-\hat{\bm{K}}_0\left\{1-\frac{v}{c}(\hat{\bm{K}}_0\cdot\hat{\bm{v}})+
	\frac{v^2}{2c^2}\left[2(\hat{\bm{K}}_0\cdot\hat{\bm{v}})^2-1\right]\right\}
	-\frac{\bm{v}}{c}\left[1-\frac{v}{2c}(\hat{\bm{K}}_0\cdot\hat{\bm{v}})\right]
	+O\left(\frac{v}{c}\right)^3\,,
\end{eqnarray}
where $\hat{\bm{v}}=\bm{v}/v$. The expression can be checked with, for example,
Equation (7.6) in~\citet{Klioner:1992}. The aberration delay can be calculated
by $\Delta_{\rm A}=\Delta\Phi/2\pi\nu$, and 
\begin{equation}
	\sin\Delta\Phi=\hat{\bm{e}}_3\cdot\frac{(-\hat{\bm{K}}_0\times\hat{\bm{e}}_3)
	\times( -\hat{\bm{K}}'_0 \times\hat{\bm{e}}_3)}{\big|
	-\hat{\bm{K}}_0\times\hat{\bm{e}}_3 \big| \cdot
	\big|-\hat{\bm{K}}'_0\times\hat{\bm{e}}_3 \big|}\,.
\end{equation}
Keep to the second order in $v/c$, the above equation then gives 
\begin{equation}
	\Delta_{\rm A}=-\frac{1}{2\pi\nu}\frac{\bm{v}\cdot(\hat{\bm{K}}_0\times
	\hat{\bm{e}}_3)}{c|\hat{\bm{K}}_0\times \hat{\bm{e}}_3|^2}\left(1+
	\frac{\bm{v}\cdot\hat{\bm{K}}_0}{2c} +\frac{\bm{v}\cdot\left[\hat{\bm{K}}_0
	-(\hat{\bm{K}}_0\cdot\hat{\bm{e}}_3) \hat{\bm{e}}_3\right]}{c|\hat{\bm{K}}_0
	\times\hat{\bm{e}}_3|^2}\right)\,.
\end{equation}
%%

%----------------------------------------------------------------------
\section{Numerical accuracy of the timing model}
\label{app:numerical accuracy of the timing model}
%----------------------------------------------------------------------

In Section~\ref{sec:timing model}, we constructed the numerical timing model as
well as the inverse timing model.  A key question is the numerical accuracy of
our treatment, as the data analysis of the pulsar timing observation requires us
to track the TOAs with millisecond-level accuracy over the whole observation
time span, which usually has a time scale of ${\cal O}(10)$\,yr. Therefore, here
we give a rough estimation of the numerical accuracy of our timing model and the
corresponding inverse timing model.

%----------------------------------------------------------------------
\begin{figure}[t]
	\begin{center}
		\includegraphics[width=17cm]{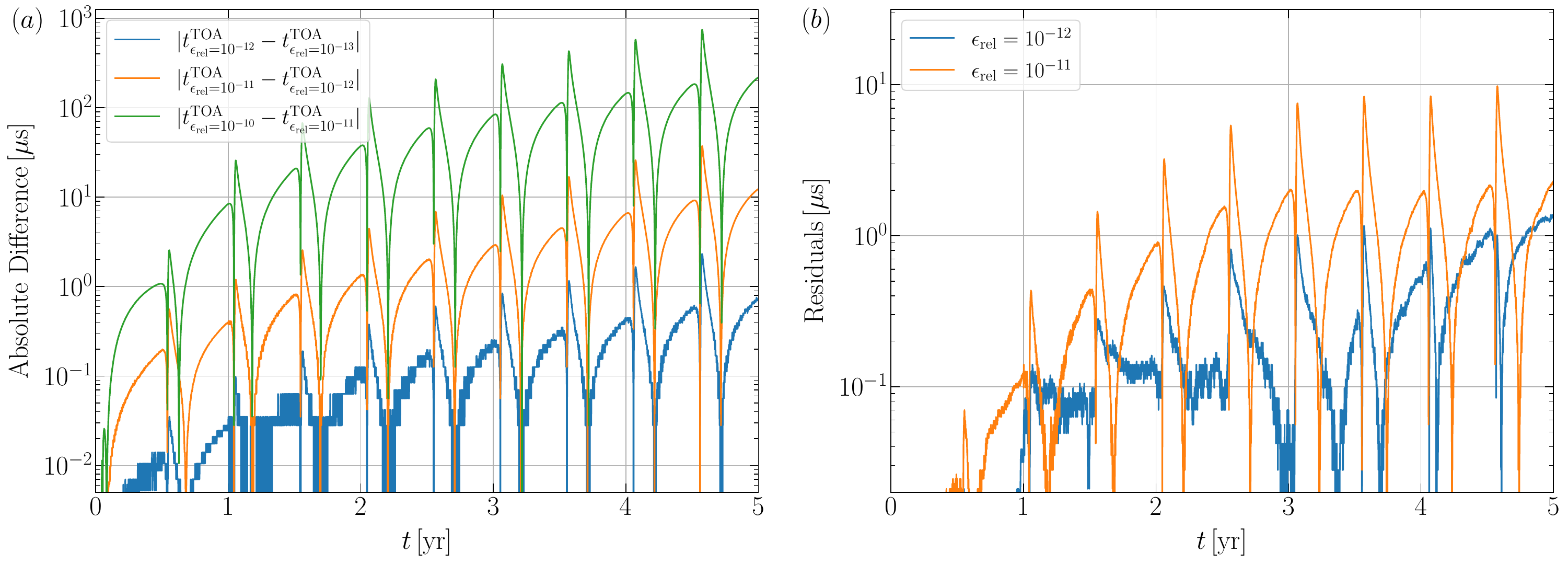}
	\caption{(a) Convergence test of  $t^{\rm TOA}$ as a function of $t$. We choose
	four different desired relative errors, namely $\epsilon_{\rm
	rel}=10^{-10}$, $10^{-11}$, $10^{-12}$, and $10^{-13}$. One can see a clear
	convergence among the results. The system parameters used for this figure
	are the same as the parameters used for Figure~\ref{fig:time delay}.
	(b) The residual caused by cumulated numerical errors in our
	integration. For a 5-yr simulation, the numerical error is at the order of
	$1\,{\rm \mu s}$ and $10\,{\rm \mu s}$ for $\epsilon_{\rm rel}=10^{-12}$ and
	$\epsilon_{\rm rel}=10^{-11}$, respectively. The error level is consistent
	with panel (a) of this figure.
	\label{fig:Convergence_test_t_toa}} 
	\end{center}
\end{figure}
%----------------------------------------------------------------------

In our code, to solve the equations of motion for the timing model or the
differential equations for the inverse timing model, we use the explicit
embedded Runge-Kutta Prince-Dormand $(8,9)$ method provided in the GSL
library~\citep{galassi2009gnu}, which has adaptive step-size control. Therefore,
the step size of the integration is mainly controlled by a parameter called the
desired relative error, $\epsilon_{\rm rel}$. In
Figure~\ref{fig:Convergence_test_t_toa}~(a), we show a convergence test for  $t^{\rm
TOA}$ as a function of $t$, which can be generated from our timing model. We
choose four $\epsilon_{\rm err}$, namely $10^{-10}$, $10^{-11}$, $10^{-12}$, and
$10^{-13}$ in our test. The results show a clear convergence. Therefore, the
numerical error for our code to generate  TOAs during a 5-yr time period can be
estimated as, for example, for $\epsilon_{\rm rel}=10^{-11}$, at the $10\,{\rm
\mu s}$ level, while for $\epsilon_{\rm rel}=10^{-12}$, at the $1\,{\rm \mu s}$
level. This error is far below the expected timing precision, so that our code
is reliable at least for the observation time span considered in this work.
Though for a longer time span, one may need to use a smaller $\epsilon_{\rm
rel}$, or even use better storage precision as performed in
Tempo2~\citep{Hobbs:2006cd}. Considering the consistency with the numerical
derivative part to be discussed in Appendix~\ref{app:numerical detail}, in this
work, we always set $\epsilon_{\rm rel}=10^{-11}$. Though we shall notice that
$\epsilon_{\rm rel}=10^{-11}$ discussed there in general leads to a smaller
integration step size due to the enlargement of the equation set. So that if one
only uses the timing model or the inverse timing model, a smaller $\epsilon_{\rm
rel}$ like $\epsilon_{\rm rel}=10^{-12}$ or $10^{-13}$ is better, but a too
small $\epsilon_{\rm rel}$ will lead to a large cumulated round-off error due to
the small integration step size.

We also investigate the numerical accuracy of the inverse timing model. We
simulate a series of TOAs with our timing model and calculate the residuals with
the true  parameters of the system and the inverse timing model. In principle,
this should give us a vanishing result. However, due to the numerical error
cumulated in the integration, a non-zero residual will exist as shown in
Figure~\ref{fig:Convergence_test_t_toa}~(b). For the 5-yr simulation, the residual caused by
numerical error is at the order of $1\,{\rm \mu s}$ and $10\,{\rm \mu s}$ for
$\epsilon_{\rm rel}=10^{-12}$ and $\epsilon_{\rm rel}=10^{-11}$, respectively.
The error level is consistent with the previous analysis and well below the
required precision limit. 

%----------------------------------------------------------------------
\section{Numerical derivatives for the Fisher matrix}
\label{app:numerical detail}
%----------------------------------------------------------------------

In Section~\ref{subsec:Fisher setup}, we introduced the Fisher matrix method
that we used to estimate the measurement precision of the system parameters. To
construct the Fisher matrix, one needs to calculate the first-order derivatives,
$\partial\mathcal{N}_i/\partial{\Theta}^{\alpha}$, as shown in
Equation~(\ref{eq:cov matrix2}). A usual treatment is to use the finite
difference method to calculate this derivative. For example, for the
second-order central difference, one has
\begin{equation}
	\frac{\partial \mathcal{N}_i}{\partial\Theta^{\alpha}}
	\approx\frac{-\mathcal{N}_i(\bar{\Theta}^\alpha+2h)
	+8\mathcal{N}_i(\bar{\Theta}^\alpha+h)-
	8\mathcal{N}_i(\bar{\Theta}^\alpha-h)+
	\mathcal{N}_i(\bar{\Theta}^\alpha-2h)}{12h}\,,
\end{equation}
where $h$ is the step size of the finite difference. The numerical error in the
above equation comes from two aspects. One is the finite difference
approximation, as one truncates the Taylor expansion at a finite order. For this
second-order formula, the error is proportional to $h^5$. Another part of the
error budget comes from the fact that the above equation is performing
subtraction for large numbers. For a 5-yr observation time span, the proper
rotation number $\mathcal{N}$ of the pulsar can reach about $5\,{\rm yr}/1\,{\rm
sec}\sim10^8$. Due to the finite precision of a stored number, this gives an
error around $10^{-8}/h$ for the {\tt double precision} we used. Therefore, to
obtain the best estimation of the derivatives, one needs to balance these two
errors. 

The above discussion can be found in standard textbooks of numerical methods.
However, we find that, for some parameters in our pulsar-SMBH systems, one
cannot find a step size $h$ that gives a satisfactory precision. In general,
this happens for angular variables that cannot be measured with high precision.
For example, the direction parameters of the SMBH's spin, $\lambda$ and $\eta$.
One can roughly think that, for these variables, changing them a lot only
introduces a small change in $\mathcal{N}_i$. Therefore, the error in the above
equation mainly comes from the finite storage precision. However, different from
those linear parameters like $\chi$ or $q$, for which one can increase the step
size to control the error, angular variables cannot have a large step size $h$
as they are periodic over $2\pi$. This leads to a large error in the Fisher
matrix calculation.

To avoid the above problem, we use a different method for calculating the
derivatives. Taking $\lambda$ as an example, one may notice that
\begin{equation}
	\frac{\partial \mathcal{N}_i}{\partial\lambda}= \big(\nu+\dot{\nu}T_i+\cdots
	\big)\frac{\partial T_i}{\partial \lambda}\,, 
\end{equation}
where $T_i=t_i-\Delta_E(t_i)$ is the corresponding emission time of the pulse.
To calculate $\partial t_i/\partial \lambda$ or $\partial \Delta_E/\partial
\lambda$, one can rely on the continuous dependence of the solutions of ordinary
differential equations on their parameters. That is, denoting the equations of
motion for pulsar timing as 
\begin{equation}
	\frac{\partial \bm{X}}{\partial t^{\rm
	TOA}}=\bm{f}\left[\bm{X}(\lambda),\lambda\right]\,,
\end{equation}
where $\bm{X}$ denotes the parameter set $\{\bm{r},\bm{v},t,\Delta_{E}\}$. One
can take a derivative of $\lambda$ for the above equation and obtain 
\begin{equation}
	\frac{\partial}{\partial t^{\rm TOA}} \frac{\partial\bm{X}}{\partial
	\lambda} =\frac{\partial\bm{f}} {\partial \bm{X}} \frac{\partial
	\bm{X}}{\partial \lambda}+ \frac{\partial \bm{f}}{\partial \lambda}\,.
\end{equation}
The above equation, together with the original equations of motion of the
pulsar, form a complete differential system that can be used to solve $\partial
t/\partial \lambda$. One can also take the derivative of the original initial
conditions to obtain the initial conditions needed here. All the derivatives
needed for constructing the above equation can be done analytically, and as a
whole, it avoids the problem of  subtraction of large numbers. Therefore, this
method can give a much accurate calculation for the derivatives we want. This
method also speeds up the calculation for the Fisher matrix a bit, as the finite
difference method requires calculating the orbital integration four times for
each parameter.

%----------------------------------------------------------------------
\begin{figure}[t]
	\begin{center}
		\includegraphics[width=17cm]{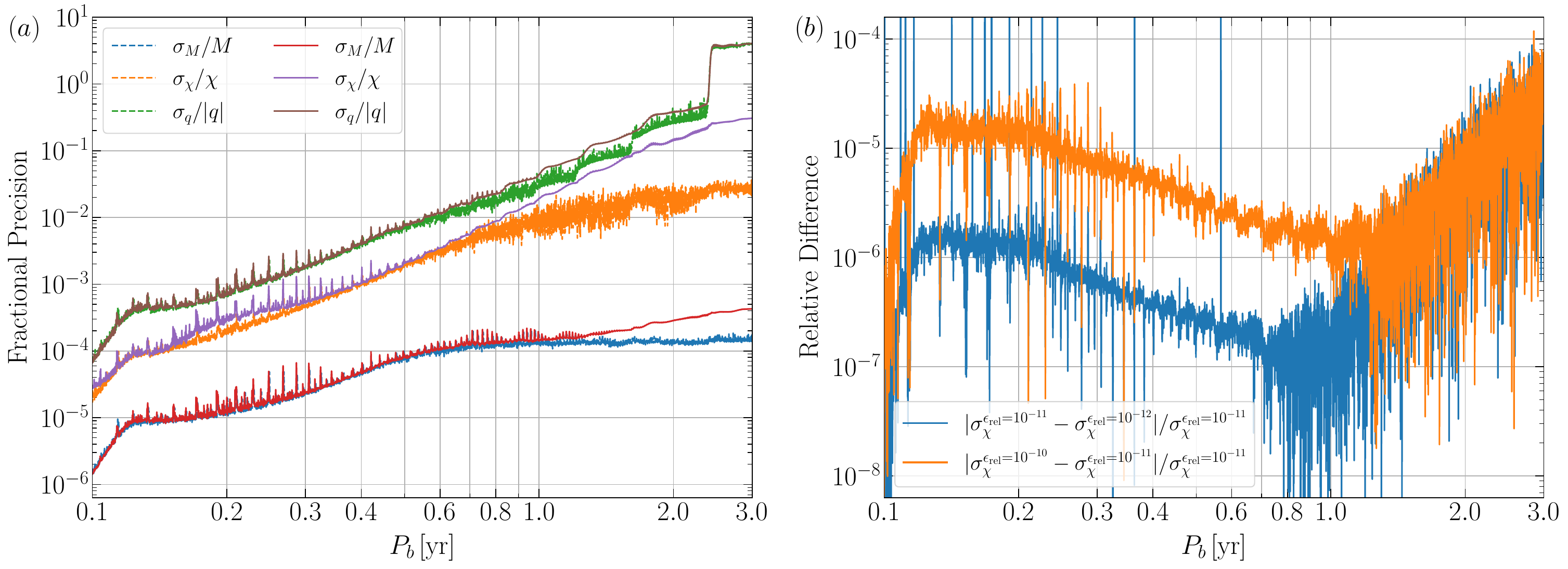}
	\caption{(a) Comparison of the parameter estimation results obtained with
	two different methods. The dashed lines employ the finite difference method,
	while the solid lines are based on solving the differential equations. One
	clearly sees that, when the pulsar's orbital periods are large, which leads
	to a worse measurement precision, the numerical errors in the results are
	largely suppressed with the method we proposed. (b) Convergence test of the
	parameter estimation results for the spin parameter $\chi$. In these
	results, we calculate the numerical derivatives using the new method
	proposed here. The desired relative errors $\epsilon_{\rm rel}$ for the
	convergence test are chosen to be $10^{-10}$, $10^{-11}$, and $10^{-12}$,
	respectively. The system parameters used for this figure are the same as the
	parameters used for Figure~\ref{fig:time delay}.
	\label{fig:Derivative_compare}} 
	\end{center}
\end{figure}
%----------------------------------------------------------------------

In Figure~\ref{fig:Derivative_compare}~(a), we show a comparison of the parameter
estimation results obtained from two methods. The dashed lines are obtained with
the finite difference method where we have fine-tuned the step size for each
parameter to obtain the best result, while the solid lines employ the method we
discussed above. All system parameters for the two cases are the same. In this
figure, for each line, we calculate 5000 points so that one can see small
structures clearly. For example, one may notice that there are a lot of spikes
in both results. These spikes are real and come from the whole integer ratio of
the pulsar's orbital period and the observation cadence (one week). That is, all
the main peaks roughly locate around $P_b=k/52\,{\rm yr}$, with $k$ an integer,
while smaller peaks allow $k$ to be a half-integer or other rational numbers.
Though for real observations, due to the more complex observation cadence, one
should not expect a pulsar-SMBH system to be located exactly in such a peak.
Similarly, around $P_b=2.5\,{\rm yr}$, one can see a step-like behavior which
comes from the fact that an additional periastron passage is observed during the
5-yr time span.

However, beside these clean structures, when the pulsar has an orbital period
$P_b\gtrsim 1\,{\rm yr}$, one can see strong oscillations in the dashed lines,
especially for the spin parameter $\chi$. This indeed comes from numerical
errors in calculating the derivatives. As discussed before, when the measurement
precision is poor, the derivative calculations for the spin direction, described
by $\lambda$ and $\eta$, are also problematic. One can see that the method
proposed here suppresses these oscillations, and we think it provides a more
reliable result. Also, as partially discussed in Section~\ref{subsec:proper
motion}, the different trends in the two sets of results are due to the correct
resolution of the proper motion effects, which are related to parameter $\Omega$
that is also a poorly constrained angular variable.

Instead of the step size $h$ for the finite difference calculation, the method
we proposed here is somehow controlled by the step size for the integration of
the differential equations. As mentioned before, in our code, we use the
explicit embedded Runge-Kutta Prince-Dormand $(8,9)$ method provided in the GSL
library~\citep{galassi2009gnu}, which has adaptive step-size control. The
parameter we can adjust is the desired relative error $\epsilon_{\rm rel}$ for
the integration. To further verify our method, by using different $\epsilon_{\rm
rel}$, we perform a convergence test of our parameter estimation results. We set
the desired relative error $\epsilon_{\rm rel}$ to  $10^{-10}$, $10^{-11}$, and
$10^{-12}$, and perform full parameter estimations.
Figure~\ref{fig:Derivative_compare}~(b) shows the relative difference of the parameter
estimation results for the spin parameter $\chi$. One can see a clear
convergence for $P_b\lesssim 1\,{\rm yr}$, while for a larger $P_b$, the
numerical error may  be dominated by the round-off error in the numerical
integration. Note that a smaller desired relative error will lead to a smaller
step size and therefore more floating-point arithmetic, so that the blue curve
has a larger part that is dominated by this error. Nevertheless, the overall
relative error in our calculation is less than $10^{-4}$, and the finite
step-size error seems to be balanced with the round-off error for $\epsilon_{\rm
rel}=10^{-11}$. Therefore, in the main text, we always use this setup. Finally,
the small relative difference also suggests that the spikes in the results
discussed before are physical and being correctly resolved, as their values are
also insensitive to the numerical setup. 

%----------------------------------------------------------------------
\begin{figure}[t]
	\begin{center}
		\includegraphics[width=8.5cm]{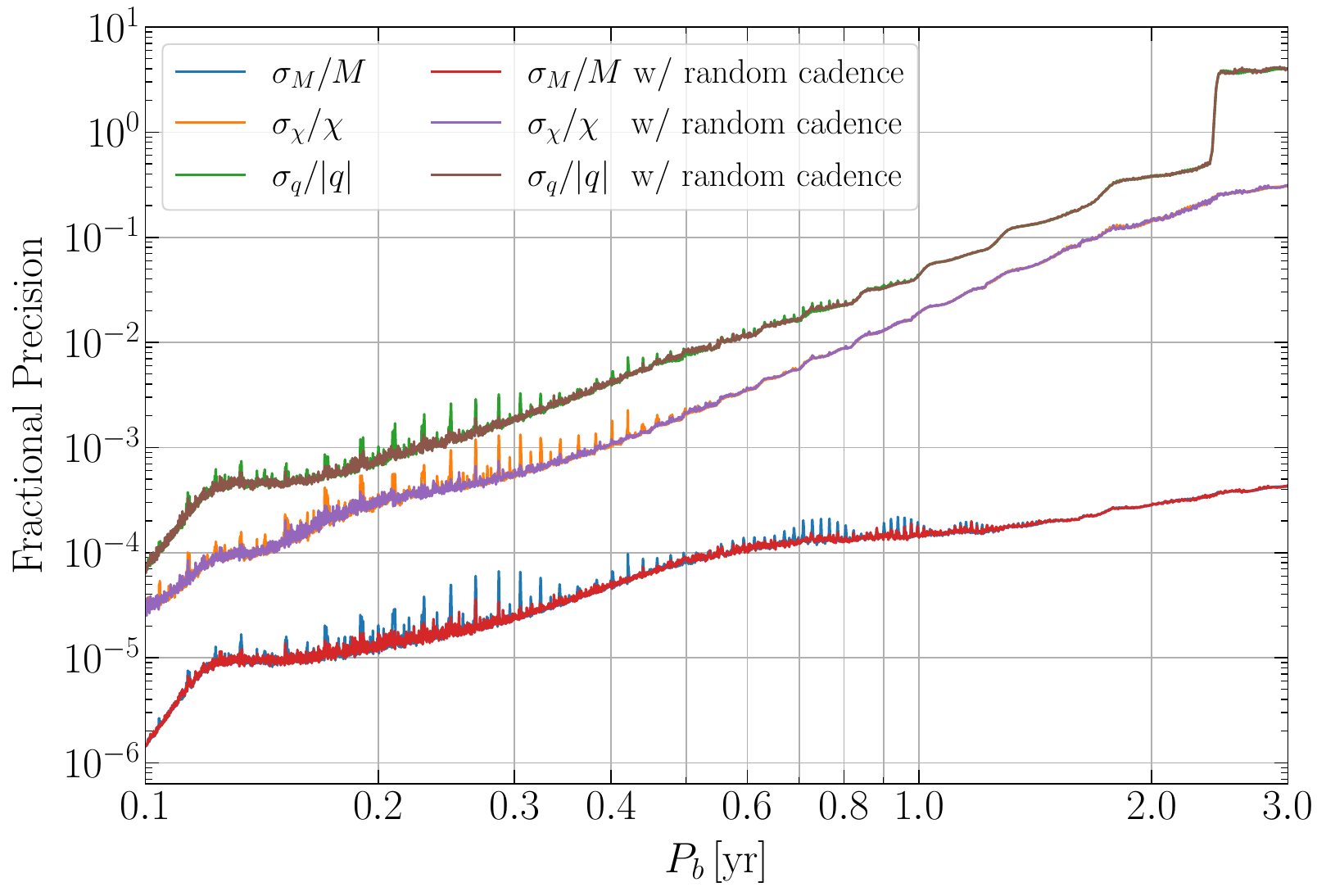}
	\caption{Comparison of the parameter estimation results with different observation cadence.}
	\label{fig:PE_ran} 
	\end{center}
\end{figure}
%----------------------------------------------------------------------

For more realistic cases, the observation cadence will not be exactly $1$ week. The randomness in the observation cadence can reduce the spike structure. In Fig.~\ref{fig:PE_ran}, we consider the observation cadence to be a uniform distribution centered around $1$ week with a range of $\pm 2$ days. As expected, with randomness in the observation cadence, the spike structure in the parameter estimation result is highly suppressed.

\bibliography{pulsarSMBH}{}
\bibliographystyle{aasjournal}

\end{document}